\documentclass[linenumbers]{aastex631}
\nolinenumbers

\usepackage{floatrow}
\newfloatcommand{capbtabbox}{table}[][\FBwidth]

\shorttitle{Seismological Studies of \textit{K2} DAVs}
\shortauthors{Hall et al.}
\graphicspath{{./}{figures/}}

\begin{document}

\title{Seismological Studies of Pulsating DA White Dwarfs Observed with the \textit{Kepler} Space Telescope and \textit{K2} Campaigns  1-8}

\author[0000-0001-6494-6098]{Weston Hall}
\affiliation{Department of Physics and Astronomy, Baylor University, Waco, TX 76798-7316}
\affiliation{Department of Physics and Astronomy, Iowa State University, Ames, IA 50011-3160}

\author{Barbara G. Castanheira}
\affiliation{Department of Physics and Astronomy, Baylor University, Waco, TX 76798-7316}

\author[0000-0002-4524-8662]{Agn\`es Bischoff-Kim}
\affiliation{Penn State Scranton, Dunmore, PA 18512}



\begin{abstract}
\nolinenumbers
 All single stars that are born with masses up to 8.5 - 10\,$M_\odot$ will end their lives as a white dwarf (WD) star. In this evolutionary stage, WDs enter the cooling sequence, where the stars radiate away their thermal energy, and are basically cooling. As these stars cool, they reach temperatures and conditions that cause the stars to pulsate. Using differential photometry to produce light curves, we can determine the observed periods of pulsation from the WD. We used the White Dwarf Evolution Code (WDEC) to calculate a grid of over one million models
 with various temperature, stellar mass and mass of helium and hydrogen layers, and calculated their theoretical pulsation periods. In this paper, we describe our approach to WD asteroseismology using WDEC models 
 and we present seismological
 studies for 29 observed DAVs in the \textit{Kepler} and \textit{K2} datasets, 25 of which have never been analyzed using these observations, and 19 of which have never been seismically analyzed in any capacity before. Learning about the internal structure of WDs place important constraints on the WD cooling sequence and our overall understanding of stellar evolution for low mass stars.

\end{abstract}



\section{Introduction} \label{sec:intro}

Variability of WD stars was discovered by chance in observations of standard stars \citep{1968ApJ...153..151L}. The advent of high-speed photometry allowed astronomers to identify periods in WDs on the order of 100 - 1200\,s, which are now known to be caused by non-radial \textit{g}-mode pulsations \citep{Fontaine_2008,doi:10.1146/annurev.astro.46.060407.145250,Althaus_2010,2019A&ARv..27....7C}

These stars are of vast importance to the field of stellar evolution, because they contain information about their previous evolutionary phases, and over 95\% of single stars will evolve into a WD \citep{1982ApJ...260..821I,1999MNRAS.302..173G}. This means that information about the interior structure of these stars can not only place constraints on the WD stars themselves, but also on the evolutionary path and structure of their progenitors. Pulsating WDs are excellent targets for asteroseismology, the only technique that can probe the interior of a star using the light coming from its photosphere, due to their simpler structure. 

The accuracy and reliability of seismological studies rely extensively on detailed and updated models of WDs, along with precision and quality of their photometric measurements. Starting in the late 1980s, astronomers established the Whole Earth Telescope (WET), in order to combat day-night aliasing \citep{1990ApJ...361..309N}. This became the premier way to take uninterrupted photometric measurements of stars, and created some of the best pictures of stellar structure for stars other than the Sun. However, despite the vast efforts of astronomers around the world, the requirements to observe faint objects like WDs ensured that only a few dozen WDs were studied by this collaboration. 
To date there are over 400 known pulsating white dwarfs with hydrogen-rich atmosphere, called DAVs or ZZ Ceti stars \citep{10.1093/mnras/stac093}.

The identification of pulsating WDs would eventually see a revolution along with the rest of the study of stellar interiors with the launch of photometric space telescopes, namely \textit{CoRoT} and \textit{Kepler}. WDs did not immediately reap these benefits at launch, as only a few were observed in the original \textit{Kepler} field. \cite{2017ApJS..232...23H} overcame that shortfall by targeting WDs to be observed in short cadence, after \textit{Kepler} was redesigned as \textit{K2} and became able to point at different fields. This survey drastically increased the number of known WDs, and provided unprecedented, high-quality measurements of pulsation periods through Fourier analysis. This study also did follow-up ground-based spectroscopic observations, as well as determination of rotation rates via asteroseismology. The quality of the observed data makes these targets suitable for an equally high-quality seismological analysis into their internal structure. 

In this work, we refine and expand ensemble asteroseismology work of the type done by \cite{Castanheira_2008,10.1111/j.1365-2966.2009.14855.x}. In those studies, core compositions were fixed to a 50:50 homogeneous mix of carbon and oxygen as an approximation to results from stellar evolution. A major difficulty of attempting to fit a large number of DAV's at once is that not all observed pulsations spectrum sample the stars the same way. Some pulsations spectra are sensitive to core structure and in that case, it is desirable to vary core parameters, or determine them through fully evolutionary models. The latter is the approach taken in works such as \cite{2012MNRAS.420.1462R,2013ApJ...779...58R,2017ApJ...851...60R}. The trade off from such an approach is coarser grids, as computing fully evolutionary sequences is computationally intensive. Also, such models rest on the assumption that the physical processes that shape the cores of white dwarfs (nuclear reaction rates, convection, core overshooting, element diffusion,...) are well understood.

A philosophically different approach is to assume that stellar evolution gives us only a broad stroke model of white dwarf interiors and that pulsations can help us reverse engineer the interior structure. Such an approach was pioneered by the works of \cite{Brassard92} and \cite{Bradley93} and gave rise to a body of work mainly focused on individual stars. One challenge of this sort of asteroseismic fitting is the choice one must make in terms of parameterization, as it is not possible to vary all of the parameters involved and we most often do not have enough observed periods to match the number of independent parameters. One choice is to vary the parameters that dictate chemical structure in the outer layers of the model (e.g. helium and hydrogen) or in the inner parts (carbon and oxygen). It is difficult to judge a priori whether a particular period spectrum is sensitive to the carbon/oxygen chemical profiles \citep{Bischoff-Kim17}, but a preliminary study of a dozen DAV period spectra showed us that for most DAV's the C/O core structure affects the fits less than the hydrogen and helium layer masses. To move forward with a unified parameterization for our pipeline fitting, we fixed the oxygen chemical profiles to that of a 0.6 solar mass fiducial model, chosen to reproduce the core structure predicted by fully evolutionary models \citep{Althaus_2010}. This is explored more for our study in Section~\ref{sec:core_sense}.

We describe here our new asteroseismic technique to analyze these stars, along with justifications for our choices in solution fitting, and we present asteroseismic results for the 29 observed DAVs in \cite{2017ApJS..232...23H}, including values for effective temperature, total mass, and hydrogen and helium mass layers. Only 10 of the 29 have been previously seismologically analyzed in some capacity, and only 4 of those used \textit{K2} observation data. In this paper, we present 19 brand new analyses of DAV stars, using high-precision photometric data from the \textit{Kepler} and \textit{K2} missions.

Other asteroseismology approaches include the Montreal group, which creates parametrized static models, searching through large parameter spaces to optimize their fits \citep{2016ApJS..223...10G,2018Natur.554...73G,2022FrASS...9.9045G}. These static models differ from other approaches as they are not derived from evolution. Similarly, the Texas group (of whom much of the work presented here was inspired by, including the use of the WDEC software), creates hot, polytropic models and cools them to specified parameters, with the ability to specify a large number of core parameter shapes and values \citep{Bischoff_Kim_2008,2018AJ....155..187B}. These approaches allow for extensive fine grids of models and a wide amount of parameters to vary. In general, this allows for high-precision fits with highly-customizable core structure, structures which are not necessarily predicted through other modelling. On the other side of WD seismology, the La Plata group specializes in using fully-evolutionary models. The benefits of this are the physical significance of the model results, however computation time is sacrificed, and they are forced to create coarser grids, and limit their parameter search \citep{2012MNRAS.420.1462R,2013ApJ...779...58R,2017ApJ...851...60R}. There also exist uncertainties in this evolution, whose impacts on asteroseismological models can be clearly established on the uncertainties in effective temperature, stellar mass, and hydrogen envelope mass \citep{2017A&A...599A..21D}.



The key difference between other DAV seismological studies and our own comes from the scope of the sample of the observed stars. When studying just one or a couple WDs, one can sample models to determine the most important parameters for the stellar interior for each star individually. Using the WDEC, this means creating extremely dense, tailored grids for each individual star, selecting the most influential parameters and trimming parameter spaces to specific ranges that best model the observed stars. In this paper, we attempt to create a uniform, structured pipeline that can be repeated and used on ensembles of DAVs, such as those presented in \cite{2017ApJS..232...23H}, in order to constrain important quantities about the star, like the mass of hydrogen and helium in the envelope. By using a standardized model grid, without customizing our calculations for each star, we must lower the amount of free parameters we examine. This is discussed more in-depth in Section~\ref{sec:models}, with a core sensitivity study done in Section~\ref{sec:core_sense}. This means that the stars examined here will probably still require more extensive, individual seismological studies. However, our goal with using this method is to prepare the future of DAV seismology to characterize seismic fits themselves along with their observed WD counterparts.

\section{Seismological Models} \label{sec:models}


The White Dwarf Evolution Code (WDEC) \citep{1975ApJ...200..306L,1990PASP..102..954W} 
has a long, rich history of development by astronomers. Originally written by Martin Schwarzchild, the code has been modified, updated, streamlined, and made available on modern machines by many astronomers.

\subsection{Input Parameters}

\cite{Bischoff_Kim_2018} contains the most recent code description, and it and its subsequent sources contain discussion to great depth on the input parameters available to change with the WDEC. These parameters affect the models in different ways and different amounts. For our study, we varied four main parameters of a WD: the total mass ($M$), the effective temperature ($T_{\textrm{eff}}$), the mass of hydrogen in the envelope ($M_\textrm{H}$) and the mass of helium in the envelope ($M_{\textrm{He}}$). As such, there are a number of parameters we chose to hold constant, because most WDs pulsate in only a few modes, which indicates a small number of observables, as suggested by \cite{Castanheira_2008} and \cite{Bischoff-Kim17}.



\begin{figure}[htbp]
    \centering
    \includegraphics[width = 0.5\linewidth]{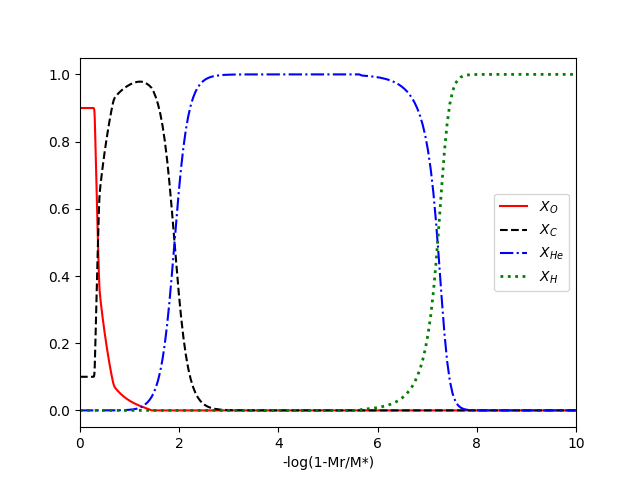}%
    \includegraphics[width = 0.5\linewidth]{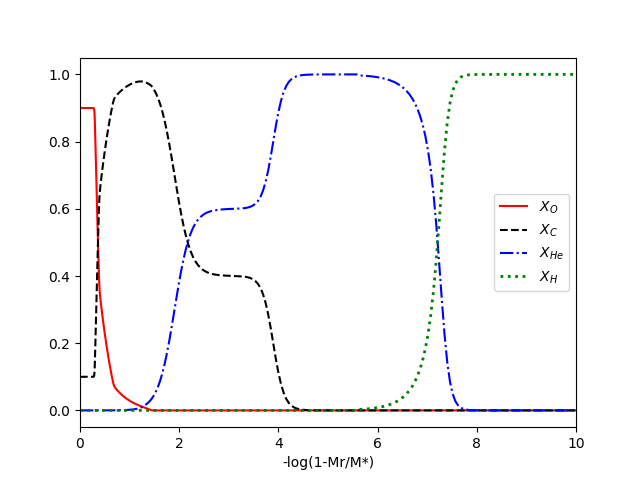}
    \caption{Chemical profiles for models with $T_{\textrm{eff}}$ = 12000\,K, $M = 0.6 M_\odot$, $\log(M_\textrm{H}/M_*)$ = -7, with $\log(M_{\textrm{He}}/M_*)$ = -2 (left) and -4 (right) respectively. The left panel shows larger He mass layer to illustrate how the He and H mass layer values affect the shape of the profiles. These different profile shapes cause different mixing in C/He transition, as shown by the flattened sections between $-\log(1-M_r/M_*)$ = 2 to 4. This will cause different modes to be trapped in the main chemical zones, as well as trapping within this transition zone.}
    \label{fig:composition_profiles}
\end{figure}

The goal in choosing our constant parameters was to create a model grid that could be used to analyze a large ensemble of WDs, that are not necessarily structured the same. As mentioned in Section~\ref{sec:intro}, with our limited choices for free parameters, we must keep the specific core parameters constant. This requires the use of a standard core profile that is more general and can represent the average WD and WDs that are close to average very well. Since nonradial pulsations are much more dependent on envelope than core, we felt it was acceptable to standardize our cores in this way, allowing for the envelope helium and hydrogen masses to change the chemical profile for each model rather than core abundances.
We fixed the parameters that dictate the shape of the core C/O profiles to best reproduce the composition profiles of \citet{1997ApJ...486..413S}, derived from stellar evolution models that evolve stars from the ZAMS and include time dependent diffusion of the elements. This profile has physical significance due to its derivation from evolutionary models, and provides a general baseline for the models we calculate. Composition profiles from two specific models calculated by WDEC for the grid, showing the transitions between different elements, specifically the C/O core and the He atmosphere, are shown in Figure~\ref{fig:composition_profiles}. We also examined the sensitivity of fits to core parameters quantitatively after creating our fitting pipeline, in Section~\ref{sec:core_sense}.

The core profile shape and composition can be parametrized with variables: $w_1$, $w_2$, $w_3$, and $w_4$, as well as $h_1$, $h_2$, and $h_3$, as described in \cite{2018phos.confE..28B}, where $w_3$ is constrained to be the difference in the size of the core and the sum of the other three free $w_n$ parameters. We also chose the helium abundance in the C/He/H region (0.60), the diffusion coefficient for He at the base of the envelope (6.0) and at the base of the pure He (9.0). These parameters are all constant through every model in the grid. We held the core parameters fixed to the values listed in Table~\ref{tab:core parameters}.

\begin{table}[htbp]
    \centering
    \caption{Values chosen for the core parameters.}
    \begin{tabular}{c c}
    \hline
    Parameter & Value \\
    \hline
        $w_1$ & 0.5 \\
         $w_2$ & 0.1\\
         $w_4$ & 0.2\\
         $h_1$ & 0.9\\
         $h_2$ & 40\% $h_1$\\
         $h_3$ & 20\% $h_1$\\
    \hline
    \end{tabular}
    \label{tab:core parameters}
\end{table}

\section{The Model Grid}

\subsection{Grid Model Variable Parameters}

We show the range and step sizes for the four parameters we varied in our asteroseismic fitting in Table~\ref{tab:grid}.
The temperature range was chosen to match that of the observed instability strip. The total mass bounds are guided by the observed mass distribution of white dwarfs \citep{Kepler17}, and the envelope mass values were chosen to represent a wide range of envelope masses, but with small enough steps to get more precise values in the solutions.

\begin{table}[htbp]
    \centering
    \caption{The parameters varied in our grid. The envelope masses are given by logarithmic fraction of the star's total mass, while the star's total mass is given in Solar masses ($M_\odot$). The total envelope mass is fixed at $10^{-2}$ $M_*$, in order to always be 2 orders of magnitude greater than the hydrogen layer.}
    \begin{tabular}{c c c c}
         \hline
         Parameter & Minimum & Maximum & Steps of  \\
         \hline
         Temperature (K) & 10600 & 12600 & 50 \\
         Total Mass ($M_\odot$) & 0.470 & 1.000 & 0.005 \\
         He envelope mass [$\log\left(M_{\textrm{He}}/M_*\right)$] & -1.50 & -4.00 & 0.25\\
         H envelope Mass [$\log\left(M_\textrm{H}/M_*\right)$] & -4.00 & -9.50 & 0.25\\
         \hline
    \end{tabular}
    \label{tab:grid}
\end{table}

For each combination of parameters, all possible periods between 100\,s and 1500\,s for all $\ell = 1$ and $\ell = 2$ modes were calculated. This would result in more than enough periods for an observed WD. Any mode higher than $\ell=2$ can usually not be observed in WD stars due to geometric cancellation in light curves \citep{Kepler_2000}. 
When referencing a specific period, we use $k$ to denote the period number, so $k = 1$ is the first (shortest) overtone, and $k = 2$ is the next shortest, and so on.

\begin{figure}[ht]
    \centering
    \includegraphics[width=\linewidth]{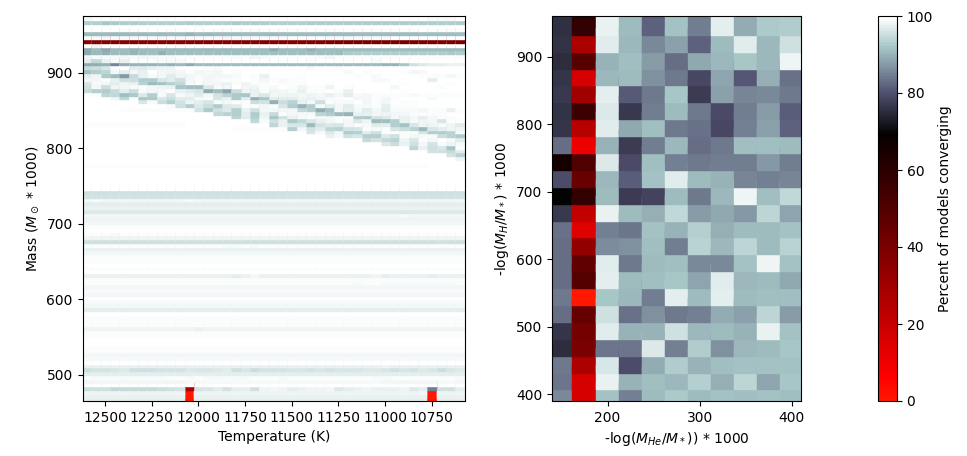}
    \caption{Heat map of the fraction of attempted models to converging models for each temperature-mass pair, and each He and H mass pair in the grid. The lighter grey areas show the finer features where models would not converge, while the redder areas show combinations where very few to no models converged, creating ``holes'' in the grid where no model exists. 97.5\% of all attempted models converged, and the ``holes'' are small enough and at favorable locations to have a very minimal effect on seismology.}
    \label{fig:converging}
\end{figure}

Not all models converged and led to the successful computation of a list of periods, though the vast majority did. Figure~\ref{fig:converging} shows the success rate of converging models in the grid per $T_{\textrm{eff}}$ and total mass combination. A total of 1,109,911 models were attempted, and 1,082,209 parameter combinations converged to form the final model grid, or 97.5\% of attempted models.  A few patterns emerge in this heat map: there are certain total masses (0.970 $M_\odot$ for example - indicating an issue with the starter model) that systematically have trouble converging, and there is an interesting band of failed models between 0.75 and 0.90 $M_\odot$. Unfortunately, the reasons for the failure of these models still remains uncertain, and a future study will need to be done to reveal the mechanisms of the code that cause a model to fail. We have certain suspicions for causes of systematic failure, such as imperfect starter models, or improper memory management by the Fortran executable when ran in sequence. Since the majority of the patterns for model failures is above the usual masses for WDs in this dataset, and therefore would not normally factor into fitting anyway, we were confident that these failed models did not affect seismology. Between 0.5 and 0.75 $M_\odot$, less than 1\% of models failed. We could have chosen to remove our high-mass models but decided to leave them in in order to convey the patterns we observed for future studies.

\subsection{Effect of the Parameters on the Periods}
The blue edge ZZ Ceti stars only pulsate with a few short periods, referred to as the first overtones. These are generally the $k = 1,2,3,4$ overtones in our models. By plotting the overtone periods vs. changing parameters of the grid, we can see how the parameters on a white dwarf model effect the calculated periods, and evaluate the physical description of what is happening in the core of ZZ Ceti stars.

\begin{figure}[ht]
     \includegraphics[width=0.5\textwidth]{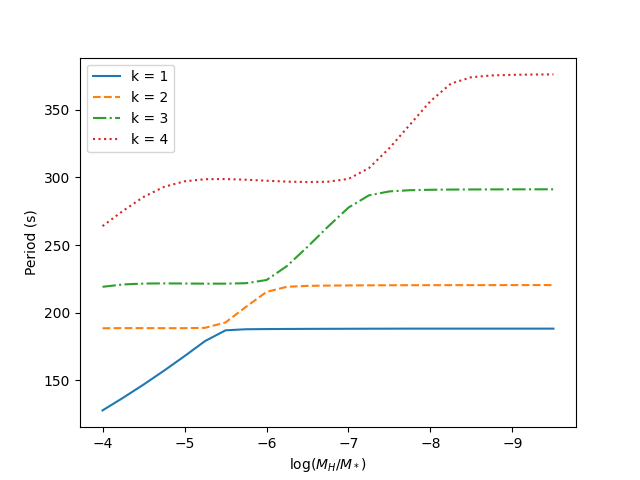}%
     \includegraphics[width=0.5\textwidth]{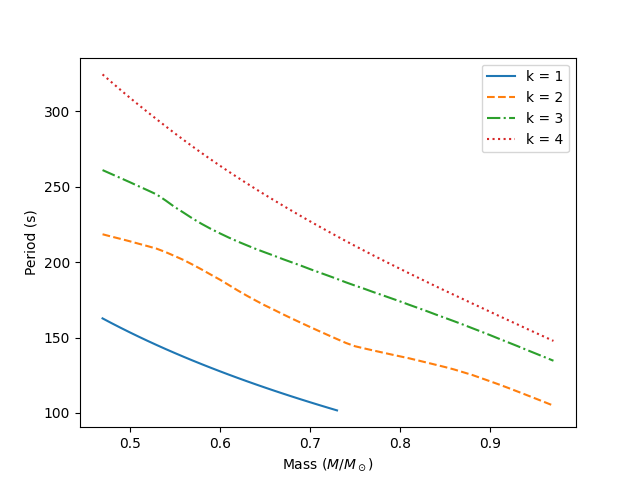}
     \includegraphics[width = 0.5\textwidth]{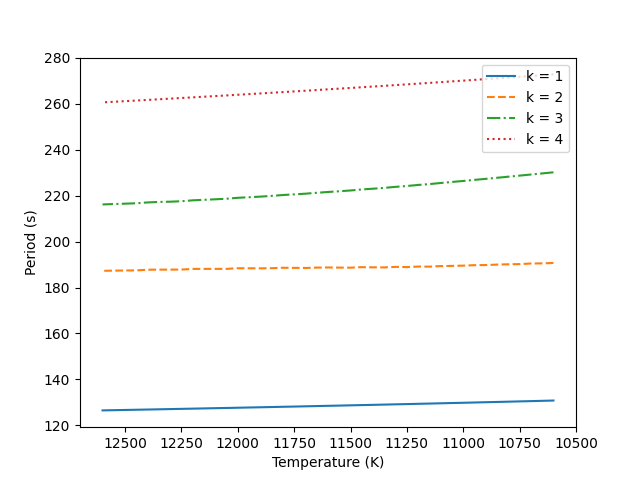}%
     \includegraphics[width = 0.5\textwidth]{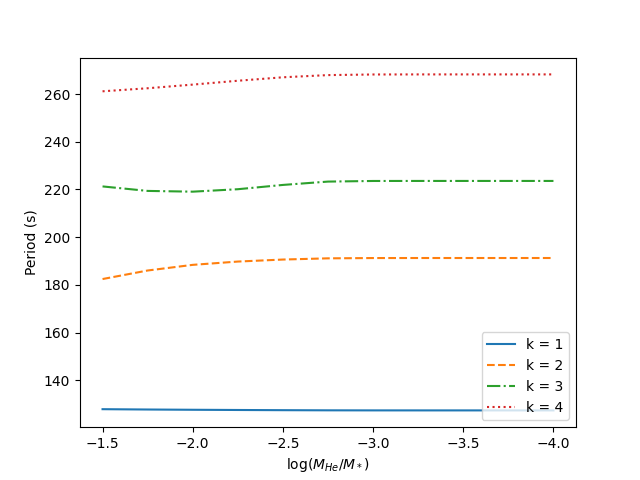}
     \caption{Upper left:  how the first four $\ell=1$ modes change vs. hydrogen mass for models of $T_{\textrm{eff}}$ = 12000\,K, $M = 0.6$ \,$M_\odot$, and $\log(M_{\textrm{He}}/M_\textrm{*}) = -2.0$. Upper right: how the first four $\ell=1$ modes change vs. total star mass for models of $T_{\textrm{eff}}=11900$ K, $\log(M_{\textrm{He}}/M_\textrm{*}) = -2.0$, and $\log(M_\textrm{H}/M_\textrm{*}) = -4.0$. Bottom left: the first four $\ell=1$ modes vs. $T_{\textrm{eff}}$ for models of $M = 0.6$\,$M_\odot$, $\log(M_{\textrm{He}}/M_\textrm{*}) = -2.0$ , and $\log(M_\textrm{H}/M_\textrm{*}) = -4.0$. Bottom right: the first four $\ell=1$ modes vs helium mass for models of $T_{\textrm{eff}}$ = 12000\,K, $M = 0.6$ \,$M_\odot$, and $\log(M_{\textrm{H}}/M_\textrm{*}) = -4.0$.}
     \label{fig:periodvhy}
\end{figure}

 Figure~\ref{fig:periodvhy} shows the change in $\ell=1$ periods for models as the varying parameters in the grid change. Decreasing the hydrogen fraction in the envelope increases the length of the periods. This shows the presence of avoided crossings in the models, which occurs when a pulsation mode takes on the properties of the next immediate $k$ mode \citep{Castanheira_2008}. This causes a variable period spacing $(\Delta P)$, depending on the thickness of the hydrogen layer. Increasing the mass of the WD significantly decreases the values of its periods, and decreases $\Delta P$. The latter means that higher mass stars have longer lists of calculated periods, making it easier for them to fit well. We must account for that effect and we describe our approach in section \ref{sec:solution fitting}.
 There is a slight increase in pulsation period as $T_{\textrm{eff}}$ is decreased. This means that as WDs cool, their modes get longer. We can physically measure the core cooling, which is the dominant cooling method for the ZZ Ceti phase. Helium mass layer has the most effect on periods at larger amounts, with little change at smaller masses.
 There is also a decrease in $\Delta P$ with increasing temperature. The combination of that effect with the dependence of $\Delta P$ on the total mass leads to the ubiquitous diagonal patterns in contour plots of best fit models (e.g. Fig. \ref{fig:KIC759 families}). The physical reason for this effect is that the period spacing depends on the average density of the model. The higher density of a high mass model can be compensated by the lower density of a hotter envelope and lead that model fitting the average spacing of an observed period spectrum, as well as a lower mass, cooler model.


\section{Solution Fitting} \label{sec:solution fitting}

The general pipeline for fitting solutions takes inspiration from the analysis of \cite{Castanheira_2008}, \cite{10.1111/j.1365-2966.2009.14855.x}, and \cite{Bischoff_Kim_2008}, with some key statistical differences later on. In order to match models to stars, we need a measure of goodness of fit. In asteroseismology, we use $S$ as our goodness of fit, where we take the sum of the squared difference in the observed vs. calculated periods, similar to a $\chi^2$ goodness of fit, as described by \cite{Castanheira_2008}, and modified for our technique in Eq.~\ref{eq:s_Val}.

\begin{equation} 
\label{eq:s_Val}
    S = \sqrt{\sum_{i=1}^{n} \frac{[P_{obs}(i)-P_{model}]^2 \times w_P (i)}{\sum_{i=1}^{n} w_P (i)}} \: \frac{N_m}{100}
\end{equation}
where $n$ is the number of observed modes, $N_m$ is the total number of $\ell$ = 1 and 2 modes calculated, and $w_P$ is the weight given to each mode. The weights are determined as proportional to the inverse of the observed period's uncertainty in order to give more weight to more precise modes, as such:
 \begin{equation}
    \label{eq:weight}
     w_P \propto \frac{1}{\sigma_{P}}
 \end{equation}
 where $\sigma_P$ is the calculated uncertainty of an observed period in seconds, as given by \cite{2017ApJS..232...23H}, and then these values for each WD are normalized between $0 < w_P \le 1$.
 
 The notable change in calculating $S$ from \cite{Castanheira_2008} is the inclusion of ${N_m}/{100}$. This was done to weight strength of fits based on how many periods were able to be calculated for a give set of parameters. A model with fewer calculated periods that closely fit an observed WD should be a higher probability solution than a model with many calculated periods that fits a larger number of stars; the inclusion of this term accounts for that. During preliminary analyses, we noted that seismology was biased towards models with low $T_{\textrm{eff}}$ and high mass. This was found to be due to cool, massive WDs having more possible periods to pulsate. By including the number of model periods in the fit, we can partially correct this bias. The factor of $1/100$ was included to normalize the $S$-values to reasonable numbers to compare with later, since the total number of periods for a WDEC model is generally between 80-120.
 
 To fit observed periods to theoretical ones, we start by matching all observed periods to the $\ell = 1$  calculated modes, which is supported by \cite{1995ApJ...438..908R}, \cite{refId0}, and \cite{2007A&A...462..989C}, since the 
 highest amplitude modes of a star are usually $\ell = 1$. However, if no solution is found for all $\ell=1$ modes, higher $\ell$ modes are tried as described. We used the mode identification done by \cite{2017ApJS..232...23H}, or through systematic incremental changes of matching individual periods to $\ell=2$. Each observed mode must only match one calculated mode in the model; this is called a one-to-one match. If multiple observed periods best fit one calculated model period, then the model is not one-to-one with the observed WD, and can be eliminated from our analysis.
 
 Once all one-to-one solutions are identified, we apply an $S$-cut, removing all models with an $S$ above a certain value. The $S$-cut is usually the number of observed periods for the WD, however if no solution has an $S$ below the number of observed periods, the $S$-cut is raised. This is why the factor of $1/100$ is included in Eq.~\ref{eq:s_Val}, in order to keep $S$ reasonably comparable to the number of observed periods.
 
 As an example of of this technique, consider \textit{KIC 7594781}. This star was first discovered in the original \textit{Kepler} dataset by \cite{2016MNRAS.457.2855G}. \textit{KIC 7594781} should be on the blue edge, and has several observed periods, with some at $\ell=1$ and $\ell=2$. Blue-edge ZZ Ceti generally have tighter constraints on their periods, and therefore their seismological fits.
 
 
 \begin{table}[htbp]
     \centering
     \begin{tabular}{c|c|c|c}
     \hline
        Period (s) & Uncertainty $\sigma$ (s) & $w_P$ & $\ell$ \\
        \hline
        206.814 & 0.00055 & 0.4545 & 1\\
        261.213 & 0.0019 & 0.1315 & 1\\
        279.647 & 0.0005 & 0.5000 & 2\\
        281.314 & 0.00080 & 0.312 & 1\\
        295.983 & 0.0011 & 0.2273 & 1\\
        328.037 & 0.00025 & 1.0000 & 2\\
        350.322 & 0.0010 & 0.25 & 2\\
        356.86 & 0.0024 & 0.1042 & 1\\
        396.146 & 0.0017 & 0.1471 & 1\\
        480.335 & 0.0052 & 0.0481 & 1\\
        683.934 & 0.014 & 0.0179 & 1\\

     \end{tabular}
     \caption{Periods of \textit{KIC 7594781} selected from \cite{2017ApJS..232...23H}. Weights follow Eq.~\ref{eq:weight}, and $\ell$'s follow those listed in \cite{2017ApJS..232...23H}, with some periods raised to $\ell=2$ when unknown in \cite{2017ApJS..232...23H}.}
     \label{tab:KIC759 periods}
 \end{table}
 
\begin{figure}
    \centering
     \begin{floatrow}
    \ffigbox{%
     \includegraphics[width = 0.47 \textwidth]{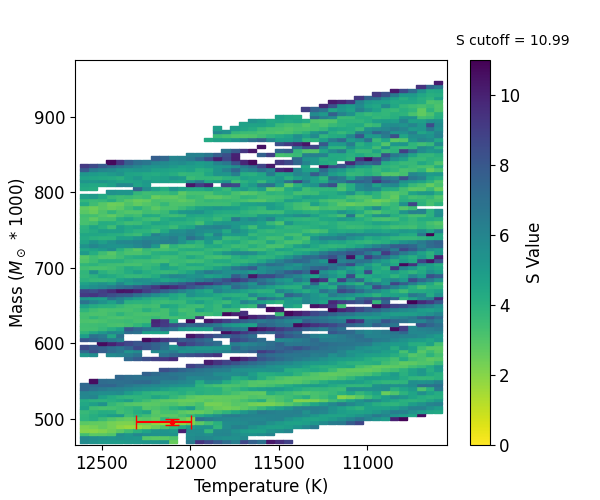}
    }{%
     \caption{All solutions for \textit{KIC 7594781} following the outline in Section~\ref{sec:solution fitting}, with the minimum $S$-value model value shown for each temperature and mass combination. Red lines mark the global minimum solution with internal uncertainty calculated using Eq.~\ref{eq:uncertainty}.}
     \label{fig:KIC759 heat map}
    }
    \ffigbox{%
      \includegraphics[width=0.53\textwidth]{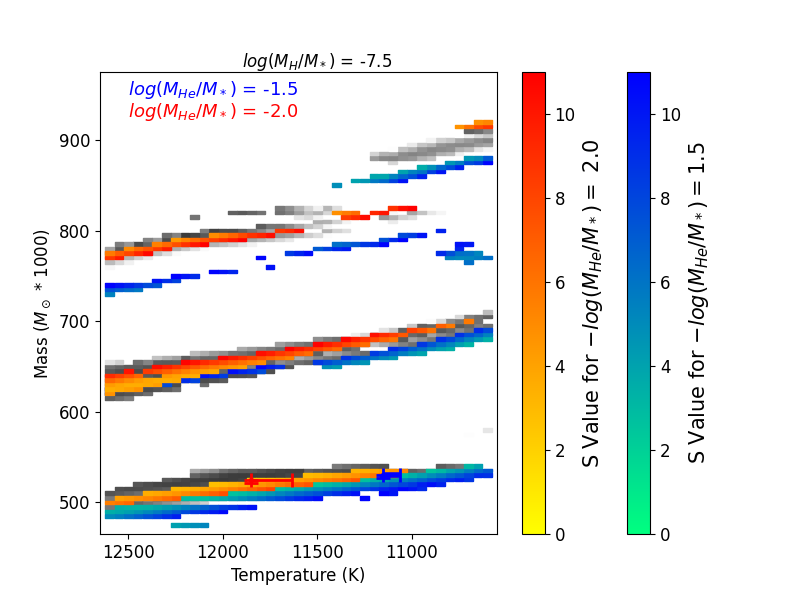}
    }{%
     \caption{Highlighted H and He masses combinations for solutions from Figure~\ref{fig:KIC759 heat map}. A slice is taken at $\log(M_\textrm{H}/M_\textrm{*}) = - 7.5$, and then $\log(M_\textrm{He}/M_\textrm{*})$ values are highlighted at $-1.5$ and $-2.5$, in blue and green respectively. Other $\log(M_\textrm{He}/M_\textrm{*})$ values are greyed out. The family minima are located at the marked solutions with respective internal uncertainties.}
     \label{fig:KIC759 families}
    }
    \end{floatrow}
 \end{figure}

 By selecting periods as $\ell=1$ and $\ell=2$, calculating an $S$ for each model, eliminating those that are not one-to-one, and then applying an $S$-cut of 11 (the number of periods listed in Table~\ref{tab:KIC759 periods}), we are left with a long list of solutions that can be visualized in the temperature-mass plane in Figure~\ref{fig:KIC759 heat map}. The global minimum is at 12100\,K and 0.495\,$M_\odot$, however it is apparent that there are other local minima, with $S$-values close to the global minima. This will be addressed in Section~\ref{sec:families}.

\subsection{Solution Uncertainty}

The uncertainty equation for asteroseismic measurements is derived in \cite{1986ApJ...305..740Z}:
\begin{equation} \label{eq:uncertainty}
    \sigma^2= \frac{d^2}{S-S_0},
\end{equation}
where $d$ is the difference between the two measurements in whatever parameter uncertainty you are calculating, $S$ is the value calculated for the model (Eq.~\ref{eq:s_Val}), and $S_0$ is the $S$-value calculated for the comparison model. The choice in selecting the two models to calculate $\sigma$ for describes the uncertainty as either internal or external for a solution. The internal uncertainty of a solution is the uncertainty between a model and its next nearest neighbor, and the external uncertainty is calculated between the minima of families of solutions.
 
 \subsection{Families of Solutions and Significant Membership} \label{sec:families}
 
 Once this list of models with a sufficiently low $S$-value are acquired, we can begin to classify solutions in this list into families, which become apparent when visualizing individual ``slices'' of Figure~\ref{fig:KIC759 heat map} at specific hydrogen/helium masses, such as Figure~\ref{fig:KIC759 families}. We begin by assigning solutions to ``families", where initially each family is composed of all solving models with the same, unique combination of hydrogen and helium masses. This process can identify up to 220 families for our grid, although usually much less actually remain after the $S$-cut. For \textit{KIC 7594781}, Figure~\ref{fig:KIC759 families} shows the family distributions of models between unique hydrogen and helium masses at $\log(M_\textrm{H}/M_\textrm{*}) = -7.5$.
 
\begin{figure}
    \centering
     \begin{floatrow}
     \capbtabbox{%
     \begin{tabular}{c|c}
     \hline
     Period (s) & $w_P$ \\
     \hline
        368.624 & 1.0000 \\
        404.603 & 0.6000 \\
        461.088 & 0.6000 \\
        505.447 & 0.5217 \\
        557.234 & 0.5217 \\
        679.474 & 0.4444 \\
        748.902 & 0.1200 \\
        799.893 & 0.1364 \\
        846.911 & 0.3000 \\
        1095.43 & 0.0632 \\
     \end{tabular} 
    }{%
      \caption{Periods of \textit{EPIC 201719578} selected from \cite{2017ApJS..232...23H}. All are $\ell=1$.}%
      \label{tab:EPIC2017 periods}
    }
    \ffigbox{%
      \includegraphics[width=0.65\textwidth]{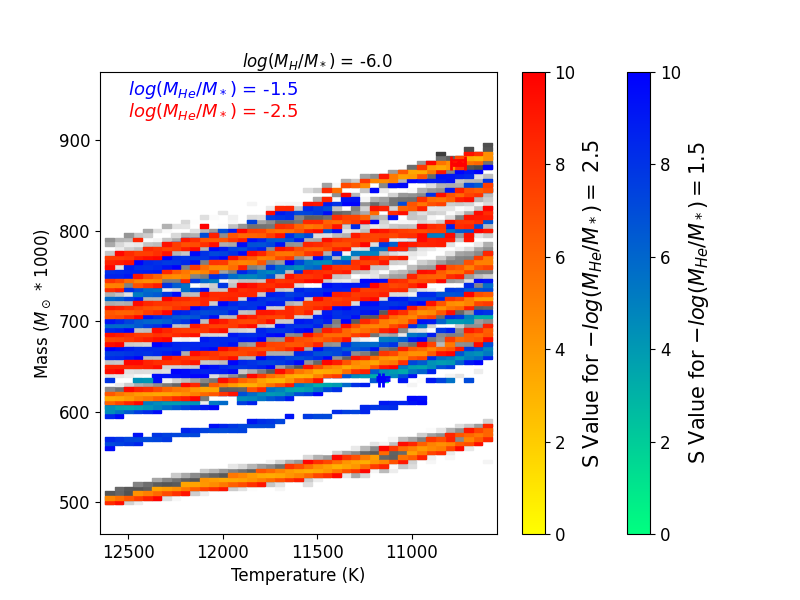}
    }{%
     \caption{Selected solution families for \textit{EPIC 201719578}, at $\log(M_\textrm{H}/M_\textrm{*})$=-6.0, with $\log(M_\textrm{He}/M_\textrm{*})$= -1.5 and -2.5 in blue and green respectively. Splitting in families of hydrogen and helium combinations are very visible into $\sim\!8$ distributions.}
     \label{fig:EPIC2017 families}
    }
    \end{floatrow}
 \end{figure}

 It can be seen that oftentimes stars will have several distributions of families on top of each other at hydrogen-helium masses combinations. For \textit{KIC 7594781}, there are about 3 or 4 possible distributions, centered at $\sim \! 0.5\,M_\odot$, $\sim\! 0.65\,M_\odot$, $\sim\!0.80\,M_\odot$, and a possible one around 0.9\,$M_\odot$. This splitting of hydrogen and helium masses combinations generally becomes even more prevalent the more periods a star has. This can be seen with the star \textit{EPIC 201719578}, whose periods are listed in Table~\ref{tab:EPIC2017 periods}, and whose seismological solution distribution at $\log(M_\textrm{H}/M_\textrm{*})=-6.0$ can be seen in Figure~\ref{fig:EPIC2017 families}, which has very distinct splitting.
 
 This presents an interesting challenge, by considering families as just unique combinations of hydrogen and helium mass values, all of these distinct distributions are considered one family, and possible well-fitting models are ignored. In order to maintain integrity of families, we have to further split hydrogen-helium combinations into these respective distributions. A naive approach would be to section off solutions into a temperature-mass grid and assign family membership based on which area they reside in, which would be subjective to the person cordoning off the grid, and varied between observed stars in the dataset. Upon close inspection, we find that solutions within a specific distribution are related via the model periods which best match the star, i.e. the \textit{k} numbers associated to the periods of a solution. By considering only the \textit{k}'s of each model, we can cluster solutions in specific hydrogen-helium masses into independent families.
 
 There are several grouping algorithms at the modern statistician's disposal, and choosing one to use for this scenario depends on a few parameters. Firstly, because of the highly distinctive and disparate behaviors of these fitting distributions from star to star, it would not be in our benefit to use a supervised learning method for ``classification''. Classification requires training on a small dataset to apply to a larger one. A supervised learning algorithm (such as \textit{k}-nearest neighbors) would more than likely overfit to one model, and not be reusable between stars, requiring more time and human analysis to assign training sets per star. Instead of classification, we decided upon ``clustering'' methods.
 
 Cluster analysis does not refer to one specific algorithm, but rather the general task, usually involving calculating distances between members, through an iterative process of knowledge discovery. These algorithms are generally unsupervised, so no training datasets are required. One of the most common ones, and the one chosen here is \textit{k}-means clustering \citep{zbMATH03340881}, which intends to partition \textit{n} observations into a specified \textit{k} clusters (\textit{k} in this case is not to be confused with the label $k$ for period numbers), where each observation belongs to a cluster with the closest mean. The drawback for our use is the dependency on selecting a \textit{k} number of clusters. By eye, one could infer 7-8 clusters from Figure~\ref{fig:EPIC2017 families}, although other hydrogen-helium combinations could differ in number. The number of clusters for each \textit{k}-means model was determined via the elbow method, which carries its own metrics, with in depth discussion in \cite{https://doi.org/10.1002/(SICI)1097-0266(199606)17:6<441::AID-SMJ819>3.0.CO;2-G}. This created a fast, efficient, and standardized way to cluster families for every star in the dataset, with relatively high accuracy.
 
\begin{figure}[htbp]
     \centering
     \includegraphics[width=0.9\textwidth]{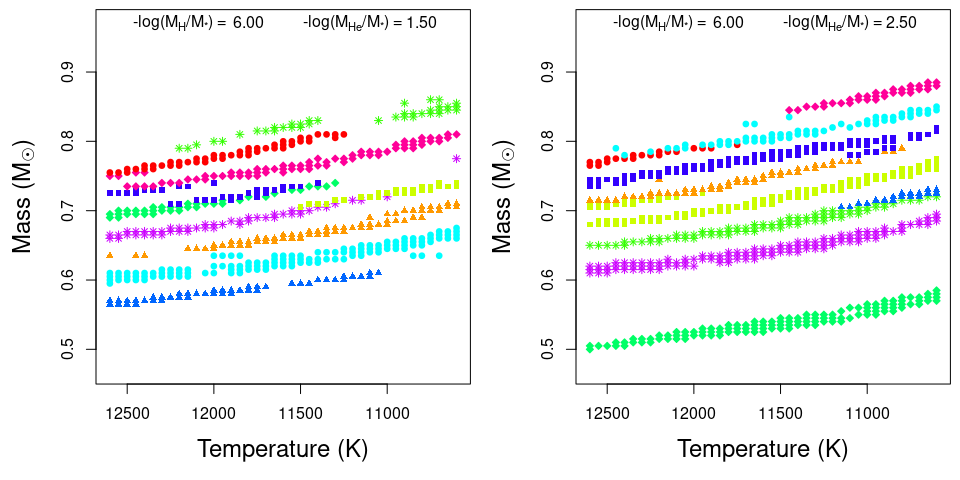}
     \caption{The identified clusters, and therefore families, for \textit{EPIC 201719578} with $\log(M_\textrm{H}/M_\textrm{*})$ = -6.0, and $\log(M_\textrm{He}/M_\textrm{*})$ = -1.5 and -2.5 as in Figure~\ref{fig:EPIC2017 families}. Color/shape of point does not correlate between the plots, and the two \textit{k}-means models were calculated independently from each other, using only the period numbers (\textit{k}'s) of the models.}
     \label{fig:clusters}
 \end{figure}

 In practice, for each hydrogen-helium mass layers pair, all solving models following the aforementioned process are selected, then only their period number \textit{k}s are clustered via \textit{k}-means. For \textit{EPIC 201719578}, a visualization can be seen in Figure~\ref{fig:clusters}, for $\log(M_\textrm{H}/M_\textrm{*})=-6.0$, where the distribution of solutions have been clustered relatively good accuracy.
 
 From here, to further narrow solution selection, we can consider family membership for the identified solutions. Certain families may contain only a handful of models, while others contain dozens, or even hundreds. These low-membership families can be considered as outliers, and their corresponding solution models can be eliminated from the solution list. This is done by only taking solutions which have ``significant membership'', where the number of models in the family must be greater than one standard deviation (1\,$\sigma$) less than the arithmetic mean number of family members for the star. This was done to strengthen confidence in the solution families that arise, and therefore the solution models derived from the families.

\subsection{Solution Selection Methods} \label{sec:selection}

Once a list of significant family solutions with the minimum-$S$ have been acquired for a star, generally dozens to hundreds of families could fit with our method. To narrow down this list, we are required to make some choices, and we want those choices to best reflect the strength of the fitted model and the accuracy of the seismological temperature and mass to other solution types, such as spectroscopy. Table~\ref{tab:solution_table} contains the absolute minimum $S$-value family solution for each star in the \textit{Kepler} and \textit{K2} dataset, however we have no inclination to prefer the global minimum family to other solution families with similarly low $S$-values, because of the massively degenerate nature of these solutions. The key is in selecting ``similarly low'' $S$-values to discriminate between viable seismological solutions and solutions that are invalidated by stronger ones.

\begin{figure}[htbp]
    \centering
    \includegraphics[width=0.9\textwidth]{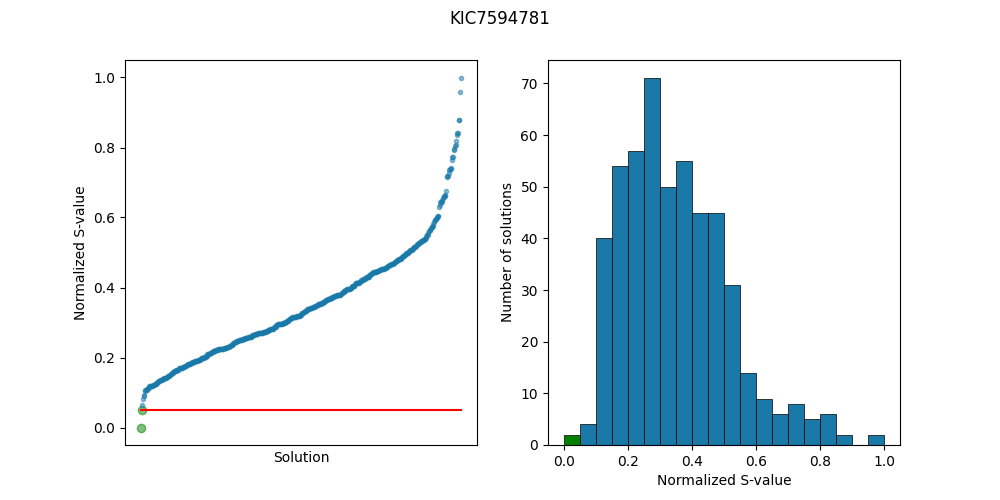}
    \includegraphics[width=0.9\textwidth]{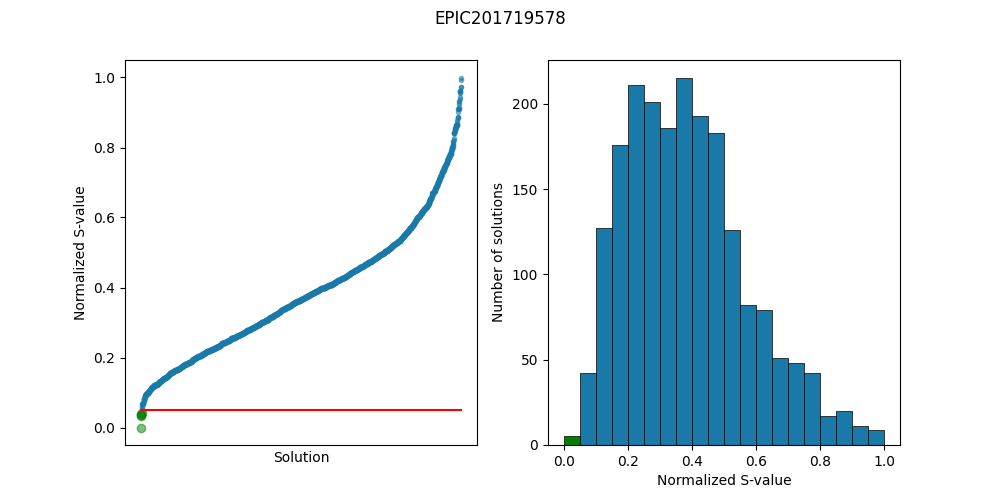}
    \caption{Solution distributions for \textit{KIC 7594781} (top) and \textit{EPIC 201719578} (bottom). The selected solutions are highlighted in green.}
    \label{fig:solution_selection}
\end{figure}

In order to trim the list of solutions down to just viable ones, we used a standardized method for each star. Taking the list of solutions obtained using hydrogen-helium combinations and machine learning clusters, we can standardize all $S$-values between 0 and 1, and then only keep those below a certain cutoff. We considered it to be better to eliminate more solutions than less, so we eliminated all solutions with a normalized $S$ above 0.05. For \textit{KIC 7594781} and \textit{EPIC 201719578}, the distribution of chosen solutions can be seen in Figure~\ref{fig:solution_selection}. By choosing those below 0.05, we can have confidence in the seismological strength of our remaining solutions to be similarly viable.

Once only viable solutions remain, we are left with our final list of solutions. Each observed star from \cite{2017ApJS..232...23H} had a varying number of solutions in their final list, ranging from one to several dozen because of the degeneracy of their observed periods. Since we are confident that the final solution list are near equivalent in seismological strength, we can use other factors to guide which solution we choose. One possible approach is turning to spectroscopy. \cite{2017ApJS..232...23H} goes into great detail of spectroscopic solutions for temperature and mass of these stars, and is the secondary method we use to choose solutions, after the minimum $S$ (Figure~\ref{tab:solution_table}). We can normalize all spectroscopic and seismological values for $T_\textrm{eff}$ and $M$ to be between 0 and 1, where 0 and 1 are the minimum and maximum values of the grid (0.47\,$M_\odot$ - 1.00\,$M_\odot$, 10600\,K - 12600\,K). From there, we can simply chose the single viable seismological solution with the lowest geometric distance to spectroscopy in the $T_\textrm{eff}$ and $M$ plane.

Creating a spectroscopy dependence in seismological solutions is not favorable for the practice of seismology, and other selection criteria exist that would not introduce as much variability in the final solutions. We plan to explore these other criteria in later studies. Another way to analyze these final valid solution lists are to approach them each as their own dataset. The benefits here are the ability to monitor constraints and precision on our parameters, like hydrogen and helium mass. The drawback is in the low amount of valid solutions that are usually retained in these lists, meaning a inherently large standard deviation for temperature and total mass, as extreme temperature and extreme mass solutions are averaged together. This is explored in our results in Section~\ref{sec:averages}.

\subsection{Core Sensitivity Study}\label{sec:core_sense}

   Because we performed asteroseismic fits of a large number of objects, we chose to focus on bulk parameters such as mass and effective temperature and envelope parameters such as the thickness of the hydrogen and helium layer mass. While these are parameters that consistently influence the best asteroseismic fits of DAVs, numerous studies have shown that g-mode pulsations in white dwarfs were sensitive to the core as well \citep[e.g.][]{Bischoff-Kim14,Giammichele18}. The WDEC is especially well suited for those types of study. Details on the parameterization of the core oxygen profile are presented in Figure 1 of \citet{Bischoff-Kim18c}. We use that parameterization in the study below.

   Modes are sensitive to the location of the chemical transitions within the model and to how sharp they are. The thickness of the helium and hydrogen layers are parameters that determine such transitions. Within the core, there is a transition from the homogeneous mix of carbon and oxygen to a region of inhomogeneous mix of carbon, oxygen, and helium. The two parameters that describe the location and the shape of that transition (sharp versus softer) are $h_1$, the central oxygen abundance and $w_1$, the location of that transition. 

   In order to quantitatively show the sensitivity of the fits to core shape, we utilized a separate grid with a coarser size (mass steps of 0.05 $M_\odot$, temperature steps of 500K, layer thickness steps of 0.5) and varying core as a supplement to our grid. 
We chose both a high and low value for these two parameters to create 4 combinations of grid parameters to examine. The numbers were picked in order to most accurately represent the total population of WDs in nature. The parameters chosen are listed in Table~\ref{tab:core parameters comp}. For the central oxygen abundance, we did not go below 0.50, consistent with the predictions of stellar evolution \citep[][e.g.]{Metcalfe02,Althaus22}. We then compared fits for \textit{KIC 7594781} and \textit{EPIC 201719578} with each of these core combinations. These two are rich pulsators with very precise fits to asteroseismology. We have used them as test stars and they are good choices for practical application of these test grids as well. 

\begin{table}[htbp]
    \centering
    \caption{Core parameters that we varied to compare with each other in the supplemental grid.}
    \begin{tabular}{c c}
    \hline
    Parameter & Value \\
    \hline
         $w_1$ & 0.1 and 0.5 \\
         $w_2$ & 0.1\\
         $w_4$ & 0.2\\
         $h_1$ & 0.5 and 0.9\\
         $h_2$ & 40\% $h_1$\\
         $h_3$ & 20\% $h_1$\\
    \hline
    \end{tabular}
    \label{tab:core parameters comp}
\end{table}

\begin{figure}
     \centering
     \includegraphics[height = 4.53 cm]{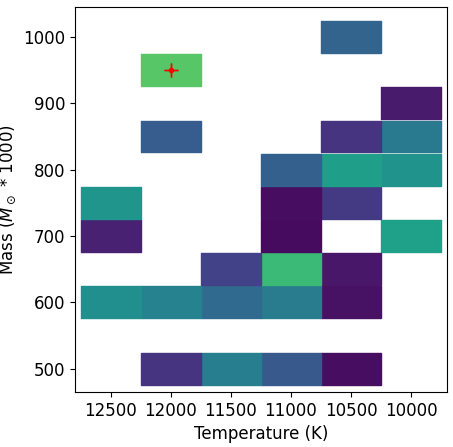}%
     \includegraphics[height = 4.53 cm]{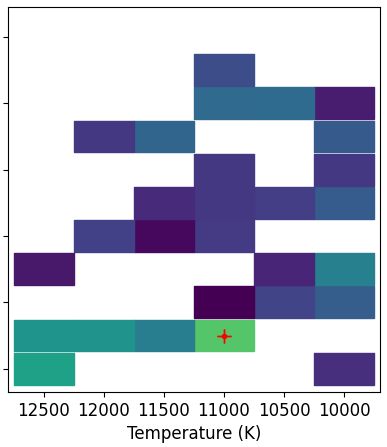}%
     \includegraphics[height = 4.47 cm]{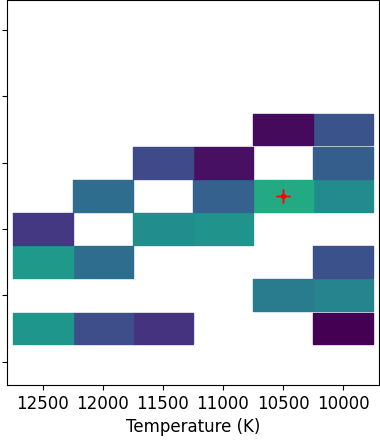}%
     \includegraphics[height = 4.53 cm]{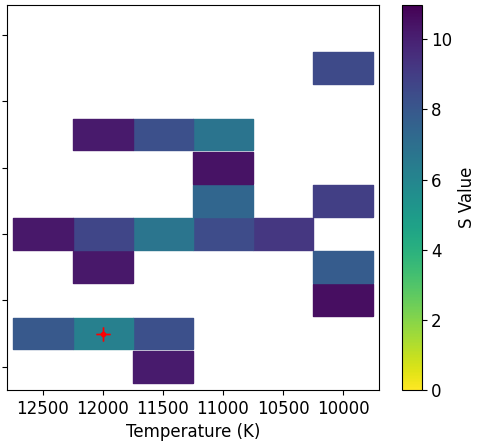}
     \caption{Contour plots for model fits to \textit{KIC 7594781} with the four combinations of core parameters in the supplemental grid. From left to right, ($h_1$, $w_1$): (0.5, 0.1), (0.5, 0.5), (0.9, 0.1), (0.9, 0.5). Individual contour plots are similar to and based on Figure~\ref{fig:KIC759 heat map}.}
     \label{fig:profile_Contours}
 \end{figure}

  We followed the same pipeline as described with the fine grid above and found model fits for each star. It is apparent that, consistent with expectations, $w_1$ is the more significant parameter to individual fit. We show the quality of fit map for \textit{KIC 7595781} in Figure~\ref{fig:profile_Contours} . Although the contours are not as refined as the main grid, we can still see that the forbidden solutions in the high-mass, hot corner of the plot are generally the same and the solution family shapes are still in the same general location. The various core solutions (along with our concluding solution from Figure~\ref{tab:average}, for comparison) are listed in Table~\ref{tab:core_comp}. Several patterns remain consistent. Except for the high mass solution of \textit{KIC 7594781} that would readily be discarded, its mass is consistent. We also recover the hydrogen and helium layer masses for the better fitting models.

For \textit{EPIC 201719578}, the core has a larger effect. We do still find higher mass solutions. But the layer masses are less consistent. It would be beneficial to refine this method for objects like \textit{EPIC 201719578}, where core structure is more significant. However, with pipeline fitting, parameters have to be selected by importance, and we selected the ones that matter most for the normal mass DAVs in the dataset.  

 \begin{table}[htbp]
     \centering
     \caption{Solutions with varying core parameters, with \textit{KIC 7594781} on top and \textit{EPIC 201719578} on bottom of the table. Solutions from Table~\ref{tab:average} are bolded at the top for reference.}
     \begin{tabular}{cllllll}
     \cline{2-7}
         & Core (h1, w1) & $T_\textrm {eff}$\,(K) & $M (M_\odot)$ & -$\log(M_\textrm{H}/M_*)$ &  -$\log(M_\textrm{He}/M_*)$  & $S$\,(s) \\[4 pt]
        \cline{2-7}
        & Average solution  & \textbf{11625 $\pm$ 475} & \textbf{0.510 $\pm$ 0.0200} & \textbf{7.38 $\pm$ 0.12} & \textbf{1.50 $\pm$ 0.00} & \\
        \cline{2-7}
        & (0.5, 0.1) &12000 & 0.950 & 6.5 & 1.5 & 2.875 \\
        \textit{KIC 7594781}&(0.5, 0.5) & 11000 & 0.550 & 7.0 & 1.5 & 2.92\\
        &(0.9, 0.1) & 12500 & 0.650 & 6.0 & 3.0 & 5.102\\
        &(0.9, 0.5) & 12000 & 0.550 & 6.5 & 3.0 & 6.257\\
     \cline{2-7}
         & Average solution & \textbf{11900 $\pm$ 514} & \textbf{0.870 $\pm$ 0.0126} & \textbf{8.20 $\pm$ 0.58} & \textbf{2.65 $\pm$ 0.73} &\\
        \cline{2-7}
        &(0.5, 0.1) & 11000 & 0.650 & 5.5 & 1.5 & 0.59  \\
         \textit{EPIC 201719578}&(0.5, 0.5) & 11500 & 0.650 & 5.5 & 3.5 & 0.714\\
        &(0.9, 0.1) & 10000 & 0.700 & 5.5 & 3.5 & 0.857\\
        &(0.9, 0.5) & 11500 & 1.000 & 4.0 & 1.5 & 0.971 \\

        \cline{2-7}
     \end{tabular}
     
     \label{tab:core_comp}
 \end{table}

 
\section{The Kepler Space Telescope}

The {\it Kepler} space telescope mission, and later {\it K2}, was designed to continuously monitor stars, for four years. The main science goal of the {\it Kepler} mission was to determine the frequency of Earth-like planets around Solar-like stars. Another scientific goal was to characterize the stars in the field. Most of the observations were in long-cadence (30\,minutes), but some targets had short-cadence (2\,min) data. After a second reaction wheel on the telescope failed, the spacecraft was no longer able to hold its observing field at a fixed position in the sky without drifting. The mission was redesigned then as {\it K2}, to scan multiple areas throughout the sky. While this would mean lower precision, the telescope was pointed to different fields, for 19 cycles, until it ran out of fuel. The short cadence observations from \textit{K2} provided a reliable way to identify DAVs in white dwarf populations. Data of stellar objects can be collected from the Mikulski Archive for Space Telescopes for NASA's {\it Kepler} \& {\it K2} Space Telescope.

\subsection{Observed DAVs}

From 2015 - 2017, observations of identified white dwarfs in the {\it K2} campaigns poured in, with correlating analysis for variability. \cite{2017ApJS..232...23H} compiled all relevant information on WDs in the {\it K2} dataset, including non-variable and variable WDs, as well as pertaining spectroscopic values for the stars in the dataset. In total, there were 29 observed pulsating WDs, all of them DAVs.

The periods listed in \cite{2017ApJS..232...23H} were calculated in different manners, namely the Linear Least-Squares (LLS) and Lorentzian (Lor) methods. Hotter DAVs have narrow pulsation peaks that can be modelled with the LLS method, while colder DAVs have longer peaks and pulsations can only be modelled through the Lor methods \citep{2017ApJS..232...23H}. For this study, we used all LLS periods when available for a star, and if not available, we used Lor modes. Lor modes have much larger uncertainties than LLS modes, sometimes by several order of magnitudes. For stars with more LLS modes than Lor, the Lor modes can be ignored when fitting solutions due to their extremely low weighting when calculating $S$.

\section{Results}

\begin{figure}[htbp]
    \centering
    \includegraphics[width=0.6\textwidth]{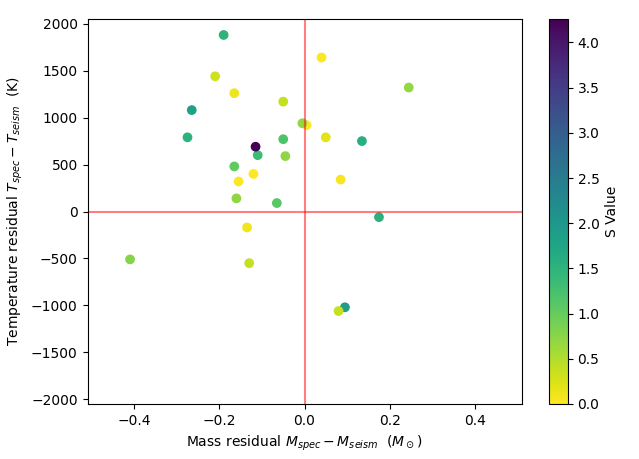}
    \caption{Residuals between spectroscopic solutions listed in \cite{2017ApJS..232...23H}, and seismological solutions listed in Table~\ref{tab:solution_table}}
    \label{fig:residuals}
\end{figure}

\begin{figure}[htbp]
    \centering
    \includegraphics[width=0.6\textwidth]{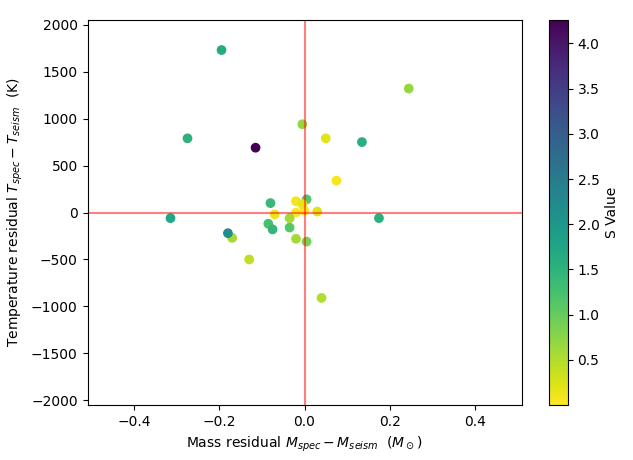}
    \caption{Residuals between spectroscopic solutions listed in \cite{2017ApJS..232...23H}, and the selected seismological solutions listed in Table~\ref{tab:selected_solution_table}}
    \label{fig:per_residuals}
\end{figure}

Table~\ref{tab:solution_table} contains the absolute minima $S$ solutions for all the stars in \cite{2017ApJS..232...23H} according to our asteroseismological study. Figure~\ref{fig:residuals} shows the residual difference between $T_\textrm{eff}$ and total mass for the seismic results here and the spectroscopic results collected in \cite{2017ApJS..232...23H}. The average seismological temperature for the WDs in this dataset is 11384\,K and the average WD mass is 0.696\,$M_\odot$. 
The minimum $S$ solutions tend to fit above the average mass for WDs (0.624\,$M_\odot$) determined via spectroscopic modeling \citep{Kepler17}.
Using the selection methods detailed in Section~\ref{sec:selection}, we choose the solutions listed in Table~\ref{tab:selected_solution_table}, with spectroscopic residual differences in Figure~\ref{fig:per_residuals}. These selections have an average seismological temperature of 11741\,K and mass of 0.667\,$M_\odot$.

\begin{longrotatetable}
\begin{deluxetable*}{p{0.15\textwidth}|p{0.07\textwidth}|p{0.07\textwidth}|p{0.05\textwidth}|p{0.05\textwidth}|p{0.05 \textwidth}|p{0.54\textwidth}}
\tablecaption{Minimum family solutions for all \textit{K2} stars. \label{tab:solution_table}}
\tablehead{
\colhead{Name} & \colhead{$T_\textrm {eff}$\,(K)} & \colhead{$M (M_\odot)$} & \colhead{$-\log(M_\textrm{H}/M_\textrm{*})$} & \colhead{ $-\log(M_\textrm{He}/M_\textrm{*})$ } & \colhead{$S$\,(s)} & \colhead{Periods ($\ell,k$)} \\}
\startdata
        \hline
        KIC4357037 & 12000 & 0.485 & 4.25 & 1.50 & 1.595 &   209.44(1,2), 261.80(1,3), 319.40(1,4), 361.29(1,5), 421.24(1,6), 530.18(1,8) \\
        KIC4552982 & 10650 & 0.715 & 4.75 & 2.75 & 0.73 &   358.21(1,6), 828.93(1,18), 866.53(1,19), 906.93(1,20), 949.91(1,21), 982.44(1,22), 1061.54(1,24), 1286.93(1,29) \\
        KIC7594781 & 12100 & 0.495 & 7.25 & 1.50 & 1.534 &   206.15(1,1), 257.60(1,2), 279.76(2,32), 257.60(1,2), 314.06(1,3), 325.89(2,33), 359.53(2,34), 361.74(1,4), 401.34(1,5), 480.10(1,6), 684.17(1,10) \\
        KIC10132702 & 11050 & 0.730 & 6.75 & 2.50 & 0.375 &   459.94(1,8), 620.14(1,12), 645.56(2,55), 725.21(2,58), 811.60(2,61), 934.12(2,66) \\
        KIC11911480 & 10800 & 0.845 & 6.50 & 3.50 & 1.829 &   167.75(1,2), 202.66(1,3), 253.83(1,4), 298.01(1,5), 325.46(1,6) \\
        EPIC60017836 & 10800 & 0.765 & 6.25 & 3.50 & 1.034 &   370.87(1,6), 843.49(1,18), 850.49(2,65), 958.10(1,21), 1096.71(2,75), 1127.41(2,76), 1189.43(1,26) \\
        EPIC201355934 & 12600 & 0.720 & 6.00 & 3.50 & 0.369 &   151.68(1,1), 175.90(1,2), 261.18(1,4) \\
        EPIC201719578 & 11300 & 0.635 & 6.00 & 1.50 & 1.103 &   362.47(1,5), 412.55(1,6), 449.41(1,7), 503.66(1,8), 541.07(1,9), 703.51(1,13), 750.67(1,14), 797.38(1,15), 843.61(1,16), 1099.10(1,22) \\
        EPIC201730811 & 12000 & 0.690 & 6.00 & 2.00 & 1.361 &   160.34(1,1), 187.03(1,2), 204.19(1,3), 274.99(1,4), 350.70(1,5) \\
        EPIC201802933 & 10650 & 0.870 & 9.00 & 3.50 & 1.504 &   134.71(1,1), 195.70(1,2), 244.34(1,3), 278.85(1,4), 299.53(1,5), 397.81(1,7) \\
        EPIC201806008 & 10800 & 0.730 & 4.50 & 3.50 & 0.0* &   412.40(1,8) \\
        EPIC206212611 & 10800 & 0.755 & 8.00 & 2.50 & 0.01 &   1042.31(1,21), 1290.31(2,26) \\
        EPIC210397465 & 10750 & 0.500 & 4.75 & 2.75 & 1.191 &   663.21(1,10), 699.79(1,11), 749.22(1,12), 976.27(1,16), 1074.63(2,18), 1233.56(2,63), 1278.91(1,22), 1389.51(1,24) \\
        EPIC211596649 & 11100 & 0.510 & 6.50 & 1.50 & 0.194 &   259.09(1,2), 295.11(1,3), 326.05(1,4) \\
        EPIC211629697 & 11400 & 0.890 & 8.50 & 3.25 & 0.797 &   491.08(1,11), 1096.47(1,27), 1142.47(1,28), 1197.72(1,29), 1238.47(1,30), 1282.02(2,92), 1306.91(1,32), 1349.67(2,95) \\
        EPIC211914185 & 12300 & 0.635 & 6.00 & 1.50 & 0.683 &   170.87(1,1), 203.67(1,2) \\
        EPIC211916160 & 10900 & 0.575 & 7.75 & 3.00 & 0.0* &   201.14(1,1) \\
        EPIC211926430 & 10950 & 0.865 & 9.00 & 3.75 & 1.519 &   116.84(2,36), 167.05(2,38), 196.58(1,2), 241.65(1,3), 283.68(1,4), 299.71(1,5) \\
        EPIC228682478 & 12000 & 0.635 & 5.75 & 3.00 & 0.036 &   192.11(1,2), 293.43(1,4), 393.26(2,44) \\
        EPIC229227292 & 12550 & 0.525 & 9.00 & 1.75 & 1.902 &   304.52(1,3), 389.21(1,4), 504.72(1,6), 1102.82(1,18), 1116.48(2,59), 1176.05(2,61), 1224.30(1,20), 1245.58(2,63), 1322.87(1,22) \\
        EPIC229228364 & 11500 & 0.755 & 7.25 & 3.25 & 0.093 &   1070.80(1,23), 1118.15(1,24), 1205.30(1,26) \\
        EPIC220204626 & 11800 & 0.870 & 9.00 & 3.75 & 0.722 &   505.72(1,11), 511.68(2,56), 580.36(1,13), 626.06(1,14), 676.94(1,15), 786.10(2,18), 798.26(1,19) \\
        EPIC220258806 & 12200 & 0.775 & 8.00 & 1.50 & 4.259 &   111.92(2,37), 139.82(1,1), 139.84(2,38), 176.34(2,40), 186.05(1,2), 241.21(1,3), 270.40(1,4), 303.14(1,5), 304.43(2,45), 347.66(2,47), 354.54(1,6) \\
        EPIC220347759 & 10600 & 0.845 & 7.00 & 1.50 & 1.397 &   130.99(1,1), 147.64(2,39), 192.43(1,3), 253.14(1,4), 277.14(1,5) \\
        EPIC220453225 & 12600 & 0.540 & 6.00 & 3.50 & 0.379 &   309.72(1,4), 670.57(1,11), 832.48(1,14), 918.54(1,16) \\
        EPIC229228478 & 11350 & 0.735 & 8.25 & 1.50 & 0.111 &   121.04(2,34), 201.00(1,2), 358.68(2,43) \\
        EPIC229228480 & 11000 & 0.680 & 4.25 & 3.75 & 0.029 &   255.68(1,4), 292.99(1,5) \\
        EPIC210377280 & 10950 & 0.575 & 8.75 & 1.50 & 0.693 &   444.26(1,5), 531.49(1,7), 628.73(1,9), 663.19(1,10), 797.80(1,12), 944.40(1,15), 995.86(1,16) \\
        EPIC220274129 & 10650 & 0.830 & 8.25 & 2.50 & 0.328 &   272.85(1,4), 347.76(1,6), 427.09(2,47), 459.94(1,8), 680.21(1,14), 736.92(1,15), 759.33(1,16), 1387.13(1,30) \\
        \hline
\enddata
\tablecomments{* denotes $S$-value significantly close to 0. }
\end{deluxetable*}
\end{longrotatetable}


\begin{longrotatetable}
\begin{deluxetable*}{p{0.15\textwidth}|p{0.07\textwidth}|p{0.07\textwidth}|p{0.05\textwidth}|p{0.05\textwidth}|p{0.05 \textwidth}|p{0.54\textwidth}}
\tablecaption{Selected solutions for \textit{K2} stars. \label{tab:selected_solution_table}}
\tablehead{
\colhead{Name} & \colhead{$T_\textrm{eff}$\,(K)} & \colhead{$M\,(M_\odot)$} & \colhead{$-\log(M_\textrm{H}/M_\textrm{*})$} & \colhead{ $-\log(M_\textrm{He}/M_\textrm{*})$ } & \colhead{$S$\,(s)} & \colhead{Periods ($\ell,k$)} \\}
\startdata
        \hline
KIC4357037 & 12000.0 & 0.485 & 4.25 & 1.50 & 1.595 & 209.44(1, 2),  261.80(1, 3),  319.40(1, 4),  361.29(1, 5),  421.24(1, 6),  530.18(1, 8) \\ 
KIC4552982 & 11100.0 & 0.665 & 4.50 & 1.50 & 1.116 & 355.25(1, 6),  828.85(1,18),  866.76(1,19),  907.28(1,20),  949.22(1,21),  982.48(1,22),  1052.84(1,24),  1279.93(1,29) \\ 
KIC7594781 & 12100.0 & 0.495 & 7.25 & 1.50 & 1.534 & 206.15(1, 1),  257.60(1, 2),  279.76(2,32),  257.60(1, 2),  314.06(1, 3),  325.89(2,33),  359.53(2,34),  361.74(1, 4),  401.34(1, 5),  480.10(1, 6),  684.17(1,10) \\ 
KIC10132702 & 12500.0 & 0.700 & 7.00 & 2.00 & 0.599 & 466.25(1, 8),  619.58(1,12),  646.33(2,58),  725.96(2,61),  813.87(2,65),  934.55(2,20) \\ 
KIC11911480 & 12150.0 & 0.750 & 8.25 & 2.75 & 0.594 & 171.75(2,37),  202.14(1, 2),  257.83(1, 3),  297.19(1, 4),  324.74(1, 5) \\ 
EPIC60017836 & 11400.0 & 0.685 & 5.75 & 3.75 & 1.254 & 353.92(1, 5),  840.62(1,17),  845.65(2,64),  958.80(1,20),  1099.63(2,74),  1126.59(2,75),  1188.97(1,25) \\ 
EPIC201355934 & 12550.0 & 0.720 & 6.00 & 3.75 & 0.425 & 151.76(1, 1),  176.00(1, 2),  260.97(1, 4) \\ 
EPIC201719578 & 11450.0 & 0.885 & 7.75 & 3.25 & 1.701 & 373.95(1, 7),  411.61(1, 8),  459.60(1,10),  510.58(1,12),  554.93(1,13),  670.40(1,16),  748.11(1,18),  801.12(1,20),  846.40(1,21),  1094.43(1,27) \\ 
EPIC201730811 & 12500.0 & 0.660 & 5.75 & 2.00 & 1.406 & 167.42(1, 1),  182.98(1, 2),  204.80(1, 3),  276.78(1, 4),  350.67(1, 5) \\ 
EPIC201802933 & 10800.0 & 0.875 & 9.25 & 3.75 & 1.658 & 133.06(1, 1),  194.17(1, 2),  241.41(1, 3),  277.32(1, 4),  308.33(1, 5),  397.96(1, 7) \\ 
EPIC201806008 & 11200.0 & 0.630 & 5.75 & 3.00 & 0.001 & 412.40(1, 6) \\ 
EPIC206212611 & 11000.0 & 0.620 & 4.25 & 3.00 & 0.082 & 1042.26(1,22),  1290.43(2,80) \\ 
EPIC210397465 & 11700.0 & 0.525 & 6.25 & 2.75 & 1.432 & 659.48(1,10),  711.17(1,11),  757.15(1,12),  977.44(1,16),  1067.06(2,18),  1234.54(2,65),  1279.30(1,22),  1385.16(1,24) \\ 
EPIC211596649 & 11100.0 & 0.510 & 6.50 & 1.50 & 0.194 & 259.09(1, 2),  295.11(1, 3),  326.05(1, 4) \\ 
EPIC211629697 & 11050.0 & 0.515 & 4.50 & 1.50 & 1.131 & 487.11(1, 7),  1097.75(1,20),  1142.59(1,21),  1193.24(1,22),  1244.80(1,23),  1283.10(2,70),  1298.13(1,24),  1349.75(2,72) \\ 
EPIC211914185 & 12300.0 & 0.635 & 6.00 & 1.50 & 0.683 & 170.87(1, 1),  203.67(1, 2) \\ 
EPIC211916160 & 11800.0 & 0.580 & 5.50 & 2.75 & 0.002 & 201.14(1, 2) \\ 
EPIC211926430 & 10950.0 & 0.865 & 9.00 & 3.75 & 1.519 & 116.84(2,36),  167.05(2,38),  196.58(1, 2),  241.65(1, 3),  283.68(1, 4),  299.71(1, 5) \\ 
EPIC228682478 & 12250.0 & 0.725 & 8.00 & 2.00 & 0.167 & 195.18(1, 2),  292.78(1, 4),  393.28(2,46) \\ 
EPIC229227292 & 11750.0 & 0.800 & 8.25 & 1.50 & 2.189 & 298.61(1, 5),  383.33(1, 7),  499.00(1,10),  1096.60(1,25),  1113.30(2,80),  1178.20(2,83),  1226.40(1,29),  1246.87(2,86),  1321.42(1,31) \\ 
EPIC229228364 & 11350.0 & 0.690 & 4.50 & 3.00 & 0.14 & 1070.97(1,25),  1118.18(1,26),  1204.53(1,28) \\ 
EPIC220204626 & 12250.0 & 0.705 & 7.25 & 2.50 & 0.869 & 507.19(1, 9),  514.32(2,51),  580.85(1,11),  628.40(1,12),  670.25(1,13),  785.15(2,62),  799.43(1,16) \\ 
EPIC220258806 & 12200.0 & 0.775 & 8.00 & 1.50 & 4.259 & 111.92(2,37),  139.82(1, 1),  139.84(2,38),  176.34(2,40),  186.05(1, 2),  241.21(1, 3),  270.40(1, 4),  303.14(1, 5),  304.43(2,45),  347.66(2,47),  354.54(1, 6) \\ 
EPIC220347759 & 10600.0 & 0.845 & 7.00 & 1.50 & 1.397 & 130.99(1, 1),  147.64(2,39),  192.43(1, 3),  253.14(1, 4),  277.14(1, 5) \\ 
EPIC220453225 & 12450.0 & 0.580 & 7.25 & 3.50 & 0.5 & 312.57(1, 4),  670.31(1,11),  832.70(1,14),  918.62(1,16) \\ 
EPIC229228478 & 12600.0 & 0.540 & 7.00 & 2.00 & 0.212 & 121.40(2,29),  201.02(1, 1),  358.67(2,37) \\ 
EPIC229228480 & 12300.0 & 0.645 & 4.25 & 2.00 & 0.052 & 255.72(1, 4),  292.99(1, 5) \\ 
EPIC210377280 & 10950.0 & 0.575 & 8.75 & 1.50 & 0.693 & 444.26(1, 5),  531.49(1, 7),  628.73(1, 9),  663.19(1,10),  797.80(1,12),  944.40(1,15),  995.86(1,16) \\ 
EPIC220274129 & 12150.0 & 0.655 & 9.00 & 2.25 & 0.599 & 278.53(1, 3),  356.43(1, 4),  422.74(2, 6),  468.37(1, 7),  654.41(1,11),  710.65(1,12),  758.79(1,13),  1387.46(1,27) \\ 
        \hline
\enddata
\end{deluxetable*}
\end{longrotatetable}


\subsection{Constraining Values with Averages and Uncertainties} \label{sec:averages}

\begin{figure}[htbp]
    \centering
    \includegraphics[width=0.72\linewidth]{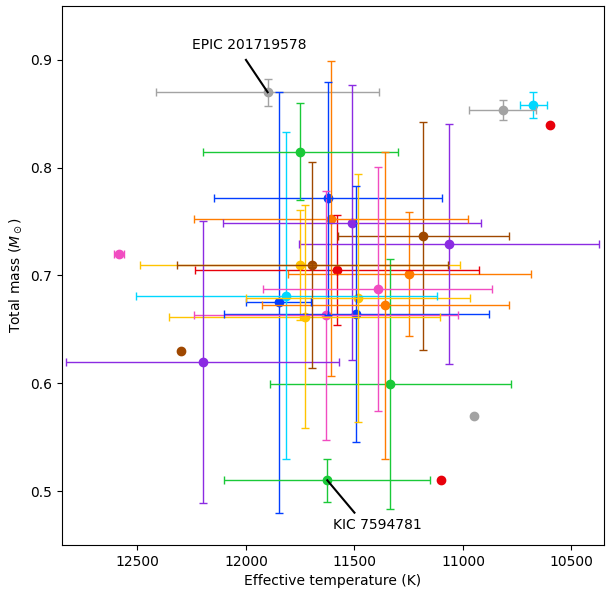}
    \caption{The average values for seismological temperature and total mass of stars in the \textit{Kepler}/\textit{K2} dataset using valid, selected solutions, with error bars demonstrating precision using standard deviation. The example stars \textit{EPIC 201719578} and \textit{KIC 7594781} are marked at their high- and low-mass solutions respectively.}
    \label{fig:k2_averages}
\end{figure}

In order to see on a more general scale what the solution fit for an observed star is, we can average the valid seismological solutions determined via Section~\ref{sec:selection}, and calculate a precision using standard deviation. Since this standard deviation calculation potentially uses multiple families, it can be understood like an external uncertainty, but not using Eq.~\ref{eq:uncertainty}. It will give us a sense of how the best seismological solutions tend to be clustered. This way, we can also gain insight into how well the hydrogen and helium layer masses are constrained. These averages and uncertainties are shown in Table~\ref{tab:average}, and plotted in Figure~\ref{fig:k2_averages}.

A very common theme was for an observed star to have only a few valid solutions, with half at a higher temperature and half at a lower temperature, with varying masses between them. This means that the standard deviations for these star's solutions can become very large. This is illustrated in Figure~\ref{fig:k2_averages} where the solutions near the center of the temperature-mass plane are averages from an equal number of extreme solutions.

Since we are performing seismology, a more useful and important factor is the precision of the hydrogen and helium thickness. Using this technique on the stars in the dataset, we can calculate a precise value for these masses, and tightly constrain them. Every star in the study is held within 2 orders of magnitude for hydrogen, with several below or around 1 order. Helium mass is held within one order of magnitude for all stars. These higher-precision hydrogen and helium masses are mostly independent of the precision for temperature and mass, which instills a lot of confidence to WD asteroseismology, since it is the only technique able to probe these values.

\begin{table}[htbp]
    \centering
    \begin{tabular}{c|c|c|c|c}
    \hline
    \colhead{Name} & \colhead{$T_\textrm{eff}$\,(K)} & \colhead{Mass ($M_\odot$)} & \colhead{$-\log(M_\textrm{H}/M_*)$} & \colhead{$-\log(M_\textrm{He}/M_*)$} \\
    \hline
KIC4357037 & 11850 $\pm$ 150 & 0.675 $\pm$ 0.1950 & 5.88 $\pm$ 1.62 & 1.75 $\pm$ 0.25 \\
KIC4552982 & 11246 $\pm$ 559 & 0.701 $\pm$ 0.0572 & 5.50 $\pm$ 1.07 & 2.36 $\pm$ 0.81 \\
KIC7594781 & 11625 $\pm$ 475 & 0.510 $\pm$ 0.0200 & 7.38 $\pm$ 0.12 & 1.50 $\pm$ 0.00 \\
KIC10132702 & 11581 $\pm$ 655 & 0.705 $\pm$ 0.0510 & 7.38 $\pm$ 1.12 & 2.62 $\pm$ 0.65 \\
KIC11911480 & 11065 $\pm$ 692 & 0.729 $\pm$ 0.1113 & 7.38 $\pm$ 1.02 & 3.20 $\pm$ 0.71 \\
EPIC60017836 & 11182 $\pm$ 392 & 0.737 $\pm$ 0.1059 & 6.68 $\pm$ 1.14 & 2.56 $\pm$ 0.73 \\
EPIC201355934 & 12583 $\pm$ 24 & 0.720 $\pm$ 0.0000 & 6.00 $\pm$ 0.00 & 3.75 $\pm$ 0.20 \\
EPIC201719578 & 11900 $\pm$ 514 & 0.870 $\pm$ 0.0126 & 8.20 $\pm$ 0.58 & 2.65 $\pm$ 0.73 \\
EPIC201730811 & 11750 $\pm$ 736 & 0.710 $\pm$ 0.0510 & 6.00 $\pm$ 0.20 & 2.08 $\pm$ 0.12 \\
EPIC201802933 & 10675 $\pm$ 63 & 0.858 $\pm$ 0.0121 & 9.00 $\pm$ 0.20 & 3.38 $\pm$ 0.24 \\
EPIC201806008 & 11492 $\pm$ 610 & 0.664 $\pm$ 0.1187 & 6.89 $\pm$ 1.61 & 2.75 $\pm$ 0.78 \\
EPIC206212611 & 11609 $\pm$ 632 & 0.753 $\pm$ 0.1461 & 6.85 $\pm$ 1.60 & 2.72 $\pm$ 0.79 \\
EPIC210397465 & 11334 $\pm$ 555 & 0.599 $\pm$ 0.1157 & 5.33 $\pm$ 0.97 & 2.95 $\pm$ 0.85 \\
EPIC211596649* & 11100 & 0.510 & 6.50 & 1.50 \\
EPIC211629697 & 11512 $\pm$ 594 & 0.749 $\pm$ 0.1278 & 7.10 $\pm$ 1.41 & 2.59 $\pm$ 0.75 \\
EPIC211914185* & 12300 & 0.630 & 6.00 & 1.50 \\
EPIC211916160 & 11631 $\pm$ 608 & 0.663 $\pm$ 0.1156 & 6.96 $\pm$ 1.53 & 2.68 $\pm$ 0.78 \\
EPIC211926430 & 10817 $\pm$ 155 & 0.853 $\pm$ 0.0094 & 8.92 $\pm$ 0.12 & 3.50 $\pm$ 0.54 \\
EPIC228682478 & 11729 $\pm$ 624 & 0.662 $\pm$ 0.1037 & 6.96 $\pm$ 1.57 & 2.68 $\pm$ 0.87 \\
EPIC229227292 & 11814 $\pm$ 693 & 0.681 $\pm$ 0.1516 & 7.68 $\pm$ 1.70 & 2.61 $\pm$ 0.97 \\
EPIC229228364 & 11622 $\pm$ 524 & 0.772 $\pm$ 0.1083 & 6.03 $\pm$ 1.52 & 2.76 $\pm$ 0.81 \\
EPIC220204626 & 11358 $\pm$ 568 & 0.672 $\pm$ 0.1425 & 8.29 $\pm$ 1.34 & 2.56 $\pm$ 0.65 \\
EPIC220258806 & 11750 $\pm$ 450 & 0.815 $\pm$ 0.0450 & 8.50 $\pm$ 0.50 & 1.50 $\pm$ 0.00 \\
EPIC220347759* & 10600 & 0.840 & 7.00 & 1.50 \\
EPIC220453225 & 12200 $\pm$ 628 & 0.620 $\pm$ 0.1307 & 6.45 $\pm$ 0.43 & 3.30 $\pm$ 0.19 \\
EPIC229228478 & 11694 $\pm$ 624 & 0.710 $\pm$ 0.0952 & 7.18 $\pm$ 1.13 & 2.23 $\pm$ 0.90 \\
EPIC229228480 & 11393 $\pm$ 528 & 0.688 $\pm$ 0.1131 & 6.14 $\pm$ 1.54 & 2.83 $\pm$ 0.81 \\
EPIC210377280* & 10950 & 0.570 & 8.75 & 1.50 \\
EPIC220274129 & 11482 $\pm$ 517 & 0.679 $\pm$ 0.1146 & 6.74 $\pm$ 1.62 & 2.74 $\pm$ 0.77 \\
    \end{tabular}
    \caption{The average valid seismological solutions.}
    \label{tab:average}
    \tablecomments{* denotes selection process eliminated all but one solution, therefore no standard deviations exist. }
\end{table}

\section{Conclusions}

We performed a systematic grid search for best fit models for the WDs listed in \cite{2017ApJS..232...23H}, between 10600 and 12600\,K and 0.47 to 1.00\,$M_\odot$, using fixed core parameters. We then described and utilized a standardized solution identification procedure to select the highest-confidence and most viable seismological model. These models provide a seismological effective temperature, total WD mass, fractional hydrogen mass, and fractional helium mass for each star in the dataset. 

We have also instilled confidence in our method of seismological analysis that we will apply to all other known WDs. Our aim is to analyze all known DAVs, and compare this technique to previous and current studies, as well as apply it to subsequent identified WDs. We are currently drafting a post factum search for both variable and non-variable WDs in the later \textit{K2} campaigns (8-19), and plan to use the same pipeline from this study for any currently unidentified variable WDs in that data. We plan to also experiment with other solution selection methods, including using parallaxes with a mass-radius relationship to narrow down the valid solution list. This would be much more empirical than spectroscopic modelling, and have less dependencies on parameters and characteristics outside of seismology's control.

    It should also be reiterated, as stated in Section~\ref{sec:core_sense}, that there are DAVs in which core structure is very influential on asteroseismic fits. The method outlined in this paper remains ineffective in determining which stars this is true for. The pipeline described above would benefit from being refined to better analyze stars like \textit{EPIC 201719578} in which core structure is significant. Expanding parameter space in pipeline fitting is computationally expensive, but such an analysis would be good to include in future studies.

In total, we analyzed 29 DAVs, using data collected from the \textit{Kepler} and \textit{K2} space telescope. Of the 29 stars, we presented 19 brand new analyses, and an additional 6 analyses of known WDs using new data. The results presented here, with emphasis on the values for hydrogen and helium layer masses, provide important values for constraining internal structure of WDs. Asteroseismology is the only technique to probe the interior of these stars, and seismological results like those from this study are directly contributing to the study of white dwarf structure and stellar evolution.

\bibliography{refs}

\begin{thebibliography}{}
\expandafter\ifx\csname natexlab\endcsname\relax\def\natexlab#1{#1}\fi
\providecommand{\url}[1]{\href{#1}{#1}}
\providecommand{\dodoi}[1]{doi:~\href{http://doi.org/#1}{\nolinkurl{#1}}}
\providecommand{\doeprint}[1]{\href{http://ascl.net/#1}{\nolinkurl{http://ascl.net/#1}}}
\providecommand{\doarXiv}[1]{\href{https://arxiv.org/abs/#1}{\nolinkurl{https://arxiv.org/abs/#1}}}

\bibitem[{{Althaus} \& {C{\'o}rsico}(2022)}]{Althaus22}
{Althaus}, L.~G., \& {C{\'o}rsico}, A.~H. 2022, \aap, 663, A167,
  \dodoi{10.1051/0004-6361/202243943}

\bibitem[{Althaus {et~al.}(2010)Althaus, C{\'{o}}rsico, Isern, \&
  Garc{\'{\i}}a-Berro}]{Althaus_2010}
Althaus, L.~G., C{\'{o}}rsico, A.~H., Isern, J., \& Garc{\'{\i}}a-Berro, E.
  2010, The Astronomy and Astrophysics Review, 18, 471,
  \dodoi{10.1007/s00159-010-0033-1}

\bibitem[{{Bischoff-Kim}(2017)}]{Bischoff-Kim17}
{Bischoff-Kim}, A. 2017, in European Physical Journal Web of Conferences, Vol.
  152, European Physical Journal Web of Conferences, 06011,
  \dodoi{10.1051/epjconf/201715206011}

\bibitem[{{Bischoff-Kim}(2018{\natexlab{a}})}]{2018phos.confE..28B}
{Bischoff-Kim}, A. 2018{\natexlab{a}}, in PHysics of Oscillating STars.
  Proceedings from the PHOST (PHysics of Oscillating STars) symposium hosted by
  the Oceanographic Observatory in Banyuls-sur-mer (France) from 2-7 September
  2018. This conference honours the life work of Professor Hiromoto Shibahashi,
  28, \dodoi{10.5281/zenodo.1715917}

\bibitem[{{Bischoff-Kim}(2018{\natexlab{b}})}]{Bischoff-Kim18c}
{Bischoff-Kim}, A. 2018{\natexlab{b}}, {Non-luminous sources of cooling in
  pulsating white dwarfs},  Zenodo, \dodoi{10.5281/zenodo.1715917}

\bibitem[{{Bischoff-Kim} \& {Montgomery}(2018)}]{2018AJ....155..187B}
{Bischoff-Kim}, A., \& {Montgomery}, M.~H. 2018, \aj, 155, 187,
  \dodoi{10.3847/1538-3881/aab70e}

\bibitem[{Bischoff-Kim \& Montgomery(2018)}]{Bischoff_Kim_2018}
Bischoff-Kim, A., \& Montgomery, M.~H. 2018, The Astronomical Journal, 155,
  187, \dodoi{10.3847/1538-3881/aab70e}

\bibitem[{{Bischoff-Kim} {et~al.}(2014){Bischoff-Kim}, {{\O}stensen}, {Hermes},
  \& {Provencal}}]{Bischoff-Kim14}
{Bischoff-Kim}, A., {{\O}stensen}, R.~H., {Hermes}, J.~J., \& {Provencal},
  J.~L. 2014, \apj, 794, 39, \dodoi{10.1088/0004-637X/794/1/39}

\bibitem[{Bischoff‐Kim {et~al.}(2008)Bischoff‐Kim, Montgomery, \&
  Winget}]{Bischoff_Kim_2008}
Bischoff‐Kim, A., Montgomery, M.~H., \& Winget, D.~E. 2008, The Astrophysical
  Journal, 675, 1505–1511, \dodoi{10.1086/527287}

\bibitem[{{Bradley} {et~al.}(1993){Bradley}, {Winget}, \& {Wood}}]{Bradley93}
{Bradley}, P.~A., {Winget}, D.~E., \& {Wood}, M.~A. 1993, \apj, 406, 661,
  \dodoi{10.1086/172477}

\bibitem[{{Brassard} {et~al.}(1992){Brassard}, {Fontaine}, {Wesemael}, \&
  {Hansen}}]{Brassard92}
{Brassard}, P., {Fontaine}, G., {Wesemael}, F., \& {Hansen}, C.~J. 1992, \apjs,
  80, 369, \dodoi{10.1086/191668}

\bibitem[{Castanheira \& Kepler(2008)}]{Castanheira_2008}
Castanheira, B.~G., \& Kepler, S.~O. 2008, Monthly Notices of the Royal
  Astronomical Society, 385, 430–444,
  \dodoi{10.1111/j.1365-2966.2008.12851.x}

\bibitem[{Castanheira \& Kepler(2009)}]{10.1111/j.1365-2966.2009.14855.x}
---. 2009, Monthly Notices of the Royal Astronomical Society, 396, 1709,
  \dodoi{10.1111/j.1365-2966.2009.14855.x}

\bibitem[{{Castanheira} {et~al.}(2007){Castanheira}, {Kepler}, {Costa},
  {Giovannini}, {Robinson}, {Winget}, {Kleinman}, {Nitta}, {Eisenstein},
  {Koester}, \& {Santos}}]{2007A&A...462..989C}
{Castanheira}, B.~G., {Kepler}, S.~O., {Costa}, A.~F.~M., {et~al.} 2007, \aap,
  462, 989, \dodoi{10.1051/0004-6361:20065886}

\bibitem[{{C{\'o}rsico} {et~al.}(2019){C{\'o}rsico}, {Althaus}, {Miller
  Bertolami}, \& {Kepler}}]{2019A&ARv..27....7C}
{C{\'o}rsico}, A.~H., {Althaus}, L.~G., {Miller Bertolami}, M.~M., \& {Kepler},
  S.~O. 2019, \aapr, 27, 7, \dodoi{10.1007/s00159-019-0118-4}

\bibitem[{{De Ger{\'o}nimo} {et~al.}(2017){De Ger{\'o}nimo}, {Althaus},
  {C{\'o}rsico}, {Romero}, \& {Kepler}}]{2017A&A...599A..21D}
{De Ger{\'o}nimo}, F.~C., {Althaus}, L.~G., {C{\'o}rsico}, A.~H., {Romero},
  A.~D., \& {Kepler}, S.~O. 2017, \aap, 599, A21,
  \dodoi{10.1051/0004-6361/201629806}

\bibitem[{Fontaine \& Brassard(2008)}]{Fontaine_2008}
Fontaine, G., \& Brassard, P. 2008, Publications of the Astronomical Society of
  the Pacific, 120, 1043, \dodoi{10.1086/592788}

\bibitem[{{Garc{\'\i}a-Berro} {et~al.}(1999){Garc{\'\i}a-Berro}, {Torres},
  {Isern}, \& {Burkert}}]{1999MNRAS.302..173G}
{Garc{\'\i}a-Berro}, E., {Torres}, S., {Isern}, J., \& {Burkert}, A. 1999,
  \mnras, 302, 173, \dodoi{10.1046/j.1365-8711.1999.02115.x}

\bibitem[{{Giammichele} {et~al.}(2022){Giammichele}, {Charpinet}, \&
  {Brassard}}]{2022FrASS...9.9045G}
{Giammichele}, N., {Charpinet}, S., \& {Brassard}, P. 2022, Frontiers in
  Astronomy and Space Sciences, 9, 879045, \dodoi{10.3389/fspas.2022.879045}

\bibitem[{{Giammichele} {et~al.}(2016){Giammichele}, {Fontaine}, {Brassard}, \&
  {Charpinet}}]{2016ApJS..223...10G}
{Giammichele}, N., {Fontaine}, G., {Brassard}, P., \& {Charpinet}, S. 2016,
  \apjs, 223, 10, \dodoi{10.3847/0067-0049/223/1/10}

\bibitem[{{Giammichele} {et~al.}(2018{\natexlab{a}}){Giammichele}, {Charpinet},
  {Fontaine}, {Brassard}, {Green}, {Van Grootel}, {Bergeron}, {Zong}, \&
  {Dupret}}]{2018Natur.554...73G}
{Giammichele}, N., {Charpinet}, S., {Fontaine}, G., {et~al.}
  2018{\natexlab{a}}, \nat, 554, 73, \dodoi{10.1038/nature25136}

\bibitem[{{Giammichele} {et~al.}(2018{\natexlab{b}}){Giammichele}, {Charpinet},
  {Fontaine}, {Brassard}, {Green}, {Van Grootel}, {Bergeron}, {Zong}, \&
  {Dupret}}]{Giammichele18}
---. 2018{\natexlab{b}}, \nat, 554, 73, \dodoi{10.1038/nature25136}

\bibitem[{{Greiss} {et~al.}(2016){Greiss}, {Hermes}, {G{\"a}nsicke}, {Steeghs},
  {Bell}, {Raddi}, {Tremblay}, {Breedt}, {Ramsay}, {Koester}, {Carter},
  {Vanderbosch}, {Winget}, \& {Winget}}]{2016MNRAS.457.2855G}
{Greiss}, S., {Hermes}, J.~J., {G{\"a}nsicke}, B.~T., {et~al.} 2016, \mnras,
  457, 2855, \dodoi{10.1093/mnras/stw053}

\bibitem[{{Hermes} {et~al.}(2017){Hermes}, {G{\"a}nsicke}, {Kawaler}, {Greiss},
  {Tremblay}, {Gentile Fusillo}, {Raddi}, {Fanale}, {Bell}, {Dennihy}, {Fuchs},
  {Dunlap}, {Clemens}, {Montgomery}, {Winget}, {Chote}, {Marsh}, \&
  {Redfield}}]{2017ApJS..232...23H}
{Hermes}, J.~J., {G{\"a}nsicke}, B.~T., {Kawaler}, S.~D., {et~al.} 2017, \apjs,
  232, 23, \dodoi{10.3847/1538-4365/aa8bb5}

\bibitem[{{Iben}(1982)}]{1982ApJ...260..821I}
{Iben}, I., J. 1982, \apj, 260, 821, \dodoi{10.1086/160301}

\bibitem[{{Kepler} {et~al.}(2017){Kepler}, {Koester}, {Romero}, {Ourique}, \&
  {Pelisoli}}]{Kepler17}
{Kepler}, S.~O., {Koester}, D., {Romero}, A.~D., {Ourique}, G., \& {Pelisoli},
  I. 2017, in Astronomical Society of the Pacific Conference Series, Vol. 509,
  20th European White Dwarf Workshop, ed. P.~E. {Tremblay}, B.~{Gaensicke}, \&
  T.~{Marsh}, 421.
\newblock \doarXiv{1610.00371}

\bibitem[{Kepler {et~al.}(2000)Kepler, Robinson, Koester, Clemens, Nather, \&
  Jiang}]{Kepler_2000}
Kepler, S.~O., Robinson, E.~L., Koester, D., {et~al.} 2000, The Astrophysical
  Journal, 539, 379, \dodoi{10.1086/309226}

\bibitem[{Ketchen \&
  Shook(1996)}]{https://doi.org/10.1002/(SICI)1097-0266(199606)17:6<441::AID-SMJ819>3.0.CO;2-G}
Ketchen, D.~J., \& Shook, C.~L. 1996, Strategic Management Journal, 17, 441,
  \dodoi{https://doi.org/10.1002/(SICI)1097-0266(199606)17:6<441::AID-SMJ819>3.0.CO;2-G}

\bibitem[{{Kotak} {et~al.}(2004){Kotak}, {van Kerkwijk, M. H.}, \& {Clemens, J.
  C.}}]{refId0}
{Kotak}, {van Kerkwijk, M. H.}, \& {Clemens, J. C.} 2004, A\&A, 413, 301,
  \dodoi{10.1051/0004-6361:20031516}

\bibitem[{{Lamb} \& {Van Horn}(1975)}]{1975ApJ...200..306L}
{Lamb}, D.~Q., \& {Van Horn}, H.~M. 1975, \apj, 200, 306,
  \dodoi{10.1086/153789}

\bibitem[{{Landolt}(1968)}]{1968ApJ...153..151L}
{Landolt}, A.~U. 1968, \apj, 153, 151, \dodoi{10.1086/149645}

\bibitem[{MacQueen(1967)}]{zbMATH03340881}
MacQueen, J. 1967, Some methods for classification and analysis of multivariate
  observations, Proc. 5th {Berkeley} {Symp}. {Math}. {Stat}. {Probab}., {Univ}.
  {Calif}. 1965/66, 1, 281-297 (1967).

\bibitem[{{Metcalfe} {et~al.}(2002){Metcalfe}, {Salaris}, \&
  {Winget}}]{Metcalfe02}
{Metcalfe}, T.~S., {Salaris}, M., \& {Winget}, D.~E. 2002, \apj, 573, 803,
  \dodoi{10.1086/340796}

\bibitem[{{Nather} {et~al.}(1990){Nather}, {Winget}, {Clemens}, {Hansen}, \&
  {Hine}}]{1990ApJ...361..309N}
{Nather}, R.~E., {Winget}, D.~E., {Clemens}, J.~C., {Hansen}, C.~J., \& {Hine},
  B.~P. 1990, \apj, 361, 309, \dodoi{10.1086/169196}

\bibitem[{{Robinson} {et~al.}(1995){Robinson}, {Mailloux}, {Zhang}, {Koester},
  {Stiening}, {Bless}, {Percival}, {Taylor}, \& {van
  Citters}}]{1995ApJ...438..908R}
{Robinson}, E.~L., {Mailloux}, T.~M., {Zhang}, E., {et~al.} 1995, \apj, 438,
  908, \dodoi{10.1086/175132}

\bibitem[{{Romero} {et~al.}(2012){Romero}, {C{\'o}rsico}, {Althaus}, {Kepler},
  {Castanheira}, \& {Miller Bertolami}}]{2012MNRAS.420.1462R}
{Romero}, A.~D., {C{\'o}rsico}, A.~H., {Althaus}, L.~G., {et~al.} 2012, \mnras,
  420, 1462, \dodoi{10.1111/j.1365-2966.2011.20134.x}

\bibitem[{{Romero} {et~al.}(2013){Romero}, {Kepler}, {C{\'o}rsico}, {Althaus},
  \& {Fraga}}]{2013ApJ...779...58R}
{Romero}, A.~D., {Kepler}, S.~O., {C{\'o}rsico}, A.~H., {Althaus}, L.~G., \&
  {Fraga}, L. 2013, \apj, 779, 58, \dodoi{10.1088/0004-637X/779/1/58}

\bibitem[{{Romero} {et~al.}(2017){Romero}, {C{\'o}rsico}, {Castanheira}, {De
  Ger{\'o}nimo}, {Kepler}, {Koester}, {Kawka}, {Althaus}, {Hermes}, {Bonato},
  \& {Gianninas}}]{2017ApJ...851...60R}
{Romero}, A.~D., {C{\'o}rsico}, A.~H., {Castanheira}, B.~G., {et~al.} 2017,
  \apj, 851, 60, \dodoi{10.3847/1538-4357/aa9899}

\bibitem[{Romero {et~al.}(2022)Romero, Kepler, Hermes, Amaral, Uzundag,
  Bognár, Bell, VanWyngarden, Baran, Pelisoli, Oliveira, Koester, Klippel,
  Fraga, Bradley, Vučković, Heintz, Reding, Kaiser, \&
  Charpinet}]{10.1093/mnras/stac093}
Romero, A.~D., Kepler, S.~O., Hermes, J.~J., {et~al.} 2022, Monthly Notices of
  the Royal Astronomical Society, 511, 1574, \dodoi{10.1093/mnras/stac093}

\bibitem[{{Salaris} {et~al.}(1997){Salaris}, {Dom{\'\i}nguez},
  {Garc{\'\i}a-Berro}, {Hernanz}, {Isern}, \&
  {Mochkovitch}}]{1997ApJ...486..413S}
{Salaris}, M., {Dom{\'\i}nguez}, I., {Garc{\'\i}a-Berro}, E., {et~al.} 1997,
  \apj, 486, 413, \dodoi{10.1086/304483}

\bibitem[{Winget \& Kepler(2008)}]{doi:10.1146/annurev.astro.46.060407.145250}
Winget, D., \& Kepler, S. 2008, Annual Review of Astronomy and Astrophysics,
  46, 157, \dodoi{10.1146/annurev.astro.46.060407.145250}

\bibitem[{{Wood}(1990)}]{1990PASP..102..954W}
{Wood}, M.~A. 1990, \pasp, 102, 954, \dodoi{10.1086/132721}

\bibitem[{{Zhang} {et~al.}(1986){Zhang}, {Robinson}, \&
  {Nather}}]{1986ApJ...305..740Z}
{Zhang}, E.~H., {Robinson}, E.~L., \& {Nather}, R.~E. 1986, \apj, 305, 740,
  \dodoi{10.1086/164288}

\end{thebibliography}
\bibliographystyle{aasjournal}

\end{document}